\documentclass[pra,twocolumn,nofootinbib]{revtex4}
\usepackage{tipa,mathrsfs}
\usepackage{bm}
\usepackage{mathrsfs}
\usepackage{amssymb}
\usepackage{amsmath}
\usepackage{amsfonts,dsfont}
\usepackage{color}
\usepackage{ulem}

\usepackage[]{graphicx}
\begin{document}
\newcommand{\half}{\frac{1}{2}}
\title{Indefinite causal order with fixed temporal order for electrons and positrons}
\author{Aur\'elien Drezet$^{1}$}
\address{$^1$Univ. Grenoble Alpes, CNRS, Grenoble INP, Institut Neel, F-38000 Grenoble, France}

\email{aurelien.drezet@neel.cnrs.fr}
\begin{abstract}
We provide an analysis of indefinite causal orders  in relativistic quantum mechanics based on the electron-positron picture of Feynman involving negative energy electrons moving backward in time. We show that genuine implementations of the paradigmatic quantum switch, but here violating some causal inequalities and with a fixed temporal order, become possible in extreme external electromagnetic field conditions allowing the presence of closed time-like curves.
\end{abstract}

\maketitle
\indent \textit{Introduction-} As it was famously stated by R.P.~Feynman~\cite{Feynman1}, the superposition principle is really at the core of quantum mechanics. Remarkably, in the last decade this principle has been  extended to the classical notion of causality leading to the definition of superposed or indefinite causal order (ICO) of quantum events~\cite{Hardy2007,Oreshkov2012,Chiribilla2012,Brukner2014}. ICOs have been intensively studied for their potential applications as resources in information processing (e. g. \cite{Araujo2014,Ebler2018,Zhao2020} for a review see \cite{Goswami2020}) and more recently quantum thermodynamics~\cite{Felce2020,Nie}. At a fundamental level ICO offer motivating perspectives for understanding the connections between quantum mechanics and general relativity as well as for their potential unification~\cite{Hardy2007,Zych2019,Dimic2020,Paunkovic2020}.\\ 
\indent Among the systems that have been proposed to illustrate the concept of ICO the quantum SWITCH (QS) is paradigmatic due to its simplicity~\cite{Chiribilla2013}. The basic idea of QS is to consider an interferometric situation where two unitary operations $\hat{A}$ and $\hat{B}$ acting on a target system with state $\Psi_0\in \mathcal{H}^{(T)}$ are applied sequencially in the order $\hat{A}$ before $\hat{B}$ or  $\hat{B}$ before $\hat{A}$ depending on the state of a control system $\phi_0$ or $\phi_1\in\mathcal{H}^{(C)}$. In the case where the control system is initially in the state $\frac{1}{\sqrt{2}}(\phi_0+\phi_1)$ the final entangled system in $\mathcal{H}^{(C)}\otimes\mathcal{H}^{(T)}$ becomes $\frac{1}{\sqrt{2}}(\phi_0\hat{A}\hat{B}+\phi_1\hat{B}\hat{A})\psi_0$ that exemplifies ICO. Moreover, after a subsequent projection on the control states $\phi_\pm=\frac{1}{\sqrt{2}}(\phi_0\pm\phi_1)$ we end up with the  states $\frac{1}{2}\phi_\pm[\hat{A},\hat{B}]_\pm\psi_0$, where  $[\hat{A},\hat{B}]_\pm=\hat{A}\hat{B}\pm\hat{B}\hat{A}$, defining pure ICOs~\cite{Chiribilla2012}. QS has been extensively studied both theoretically and experimentally~\cite{Procopio2015,Rubino2017,Goswami2018,Rubino2022} but some controversies remain about the role of time versus cause in its physical implementation. Indeed, the ideal process sketched in Fig.~1(a) relies on a two-events sequence $\mathcal{A}\prec$ (causes) $\mathcal{B}$ or $\mathcal{B}\prec\mathcal{A}$. Moreover, its physical realization involves four space-time regions  and not two as sketched in Fig.~1(b). Fundamentally this implementation of QS is not different from the `unfolded' configuration of Fig.~1(c) that is just a standard interferometer with four different unitaries ($\hat{A}$ being equivalent to $\hat{A'}$ and $\hat{B}$ to $\hat{B'}$) and for this reasons the recent tabletop implementations of QS have been criticized~\cite{MacLean2017}. However, it has been stressed that any attempt to operationally distinguish the times of interaction $t_{\mathcal{A}}$ from  $t_{\mathcal{A}'}$ or $t_{\mathcal{B}}$ from $t_{\mathcal{B}'}$ would break the coherence needed for observing ICO by providing a `which-order' information~\cite{Procopio2015,Zych2019,Rubino2022} and altering the `time-delocalization'~\cite{Oreshkov2019}. Moreover, it has also been speculated than a superposition of two quantized states of the gravitational field could lead to a genuine version of the QS involving two localized space-time events~\cite{Paunkovic2020}.\\  
\begin{figure}[h]
    \centering
    \includegraphics[width=0.8\linewidth]{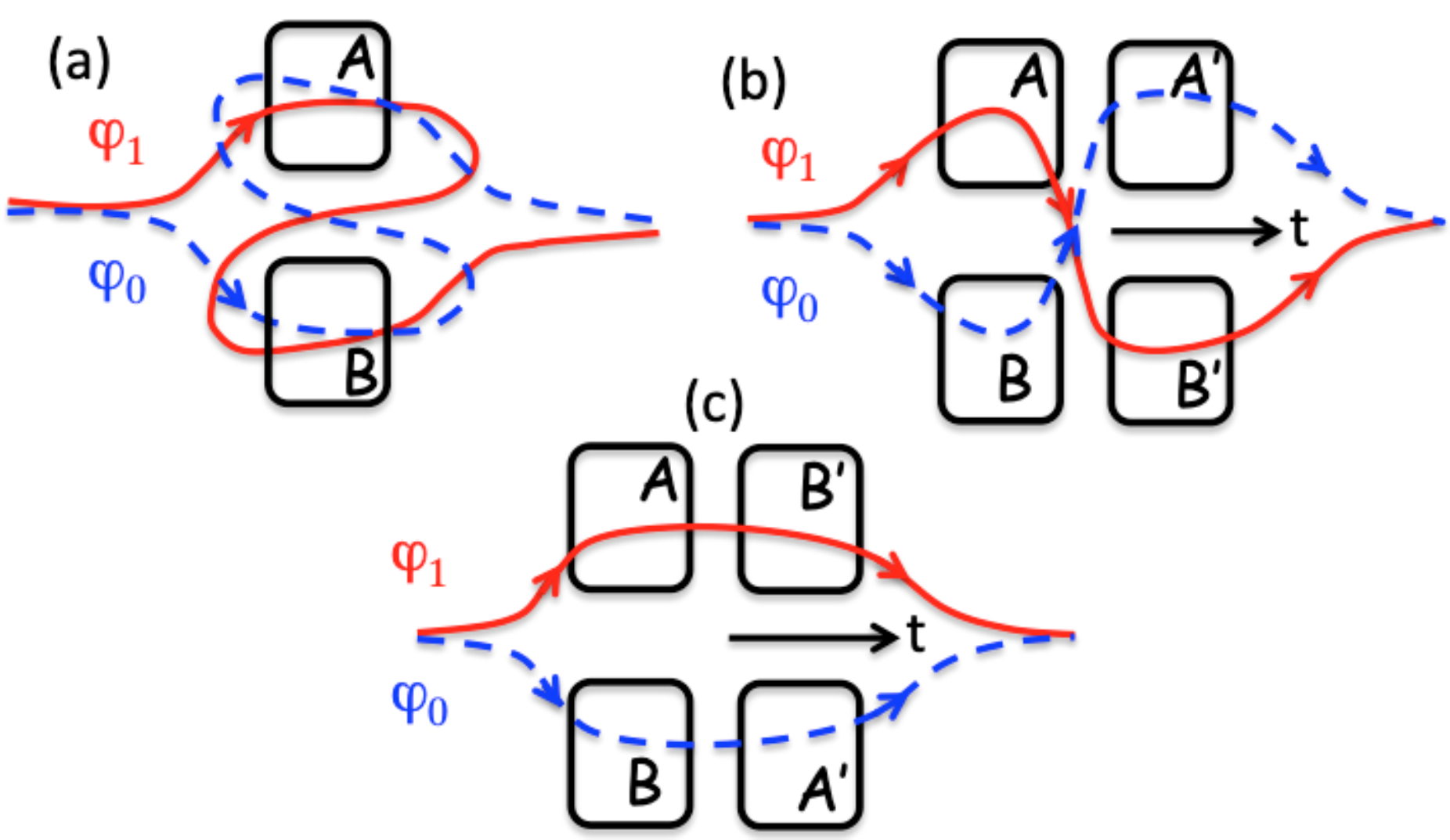}
    \caption{(a) illustrates QS with the two control states $\phi_{0,1}$ routing the target state to the unitaries $\hat{A}$ and $\hat{B}$ in different orders. (b) sketches the actual tabletop implementation involving four events in space-time at $\mathcal{A},\mathcal{A}'$ and $\mathcal{B},\mathcal{B}'$ whereas (c) is the phyically equivalent unfolded description of this QS.  }
    \label{fig:coordinate}
\end{figure}
\indent The goal of this work is to provide a different strategy relying on quantum field theory and more precisely on Feynman's theory of electrons and positrons in which an antiparticle is considered as a particle going backward in time~\cite{Feynman1949,Feynman1949b}. Importantly, this solution to the controversy doesn't require a curved spacetime but only a flat Minkowski's one. Moreover, as we show this is equivalent to demonstrate some form of retrocausality in quantum mechanics. We show that this can still be done safely in the context of quantum field theory without violating a no-signaling theorem that would  otherwise enter in conflict with microcausality and special relativity. In turn we show that our finding allows us to  violate causal inequalities~\cite{Oreshkov2012,Branciard2016} that are usually considered as being satisfied for QS and more generally with quantum mechanics~\cite{Oreshkov2012,Arujo2015,Purves2021,Dourdent2021}. Whereas this looks at first contradictory we show how a consistent picture emerges out and leads to a better understanding of causal relations in relativistic quantum mechanics.\\
\indent\textit{Results-} As it is known, the modern story of quantum electrodynamics (QED) started with the Dirac electron-hole picture for interpreting the negative energy solutions of the relativistic Dirac electron equation. Motivated by Pauli's exclusion principle Dirac introduced a bottomless energy sea with all negative energy states occupied for justifying the non observation of negative energy waves~\cite{Dirac1930}. Moreover, Feynman subsequently developped a over-all space-time view where he justified the absence of negative energy waves by a acute causal analysis of relativistic Green's functions and propagators used in scattering processes ~\cite{Feynman1949,Feynman1949b}. For instance, writting $S(x-x')$ the Green function solution of $(i\gamma^\mu\partial_\mu-m)S(x-x')=\delta^4(x-x')$, Feynman imposed the solution \begin{eqnarray}
iS_F(x-x')=\sum_{n,E_n>0}w_n(x)\bar{w}_n(x')& \textrm{for} &t<t' \nonumber\\
iS_F(x-x')=-\sum_{n,E_n<0}w_n(x)\bar{w}_n(x')&\textrm{for}& t>t'
\end{eqnarray}where $w_n(x)=w_n(\mathbf{x})e^{-iE_nt}$ are usual bispinors plane wave solutions of Dirac's equation $(i\gamma^\mu\partial_\mu-m)\psi(x)=0$ in vacuum ($\gamma^\mu$ are the standard $4\times 4$ Dirac's matrices).  Compared to the `natural' retarded Green function $iS_{ret.}(x-x')=\sum_{n}w_n(x)\bar{w}_n(x')\theta(t-t')$ Feynman's choice doesn't propagate negative energy waves ($E_n<0$) into the future direction but only positive one ($E_n>0$) therefore prohibiting unsuitable scattering to negative energy levels. The price to pay is the presence of negative energy waves scattered into the past direction. This leads to the interpretation of negative energy electron waves  going backward in time as  positive energy positron waves going forward in time.\\
\indent  Feynman's method is extensivelely used in introductory textbooks on QED where we generally emphasize the intuitive character of the results but let a rigorous justification of the rules to a more elaborated formalism based on quantum fields and second quantization (compare for example \cite{Bjorken1} and \cite{Bjorken2}). It is often overlooked that Feynman aim was not to substitute a one electron picture (with a single electron making multiple zigzags in time) to the electron-hole theory. Better, he justified rigororously his approach using the scattering matrix formalism within Dirac's hole theory (see the appendix in \cite{Feynman1949}). Furthermore, the physical consequences of taking $S_F(x-x')$ has been generally underestimated and the implications on causality mostly neglected.\\
\indent To illustrate the problem consider Feynman's second order diagram shown in Fig.~2(a) where an electron of electric charge $e$ is scattered two times by an external electromagnetic potential $V^\mu(x)$ located in two space-time regions $\mathcal{A}$ and $\mathcal{B}$. In a first quantized picture   the positive energy initial  wave $\psi_0(x)$ evolves as $\psi(x)=\psi_0(x)+ \psi^{(1)}(x)+\psi^{(2)}(x)+...$ where the second order term reads $\psi^{(2)}(x)=\int_{\Omega_{\mathcal{A}}}\int_{\Omega_{\mathcal{B}}} d^4x_{\mathcal{A}}d^4x_{\mathcal{B}}\chi(x,x_{\mathcal{A}}, x_{\mathcal{B}})+...$ with integrations over hypervolumes $\Omega_{\mathcal{A,B}}$~\cite{remark} such that
 \begin{eqnarray}
 \chi= e^2S_F(x-x_{\mathcal{A}})\textsuperimposetilde{v}(x_{\mathcal{A}})S_F(x_{\mathcal{A}}-x_{\mathcal{B}})\textsuperimposetilde{v}(x_{\mathcal{B}})\psi_0(x_{\mathcal{B}})  \label{eq2}
 \end{eqnarray}
 and $\textsuperimposetilde{v}(y):=\gamma_\mu V^\mu(y)$. Very often processes like Eq.~\ref{eq2} are considered as virtual  but this has not necessarily to be so as it was emphasized by Feynman~\cite{Feynman1962}. This holds true because $S_F(x-x')=-(i\gamma^\mu\partial_\mu+m)D_F(x-x')$ where $D_F(x-x')$, Feynman's scalar Green function solution of $(\partial^2+m^2)D_F(x-x')=\delta^4(x-x')$, is given by $\Delta_F(x)=\frac{\delta(x^2)}{4\pi}-\frac{m}{8\pi s}H_1^{(2)}(ms)$ (with $s=\sqrt{t^2-\mathbf{x}^2}$ inside the light-cone and $s=-i\sqrt{\mathbf{x}^2-t^2}$ outside). Outside the light-cone $D_F(x)$ dies off exponentially  but inside it is dominated by propagative waves associated with the Hankel function $H_1^{(2)}(ms)$~\cite{Feynman1962} that can have a physical effect even far away from the interaction zones. Moreover, in the case shown in Fig.~\ref{fig:2}(a) the two space-time volumes $\Omega_{\mathcal{A,B}}$ are disjoints and if $\Omega_{\mathcal{A}}$ is in the absolute past of $\Omega_{\mathcal{B}}$ (i.e., inside its past light-cone) then we have a retrocausal influence from $\mathcal{B}$ to $\mathcal{A}$ driven by negative energy waves propagating backward in time. Apriori, this could be used to send a signal into the past since the field $V^\mu(x_{\mathcal{B}})$ could be monitored after the interaction at $x_{\mathcal{A}}$ occurred.\\ 
\begin{figure}[h]
    \centering
    \includegraphics[width=0.8\linewidth]{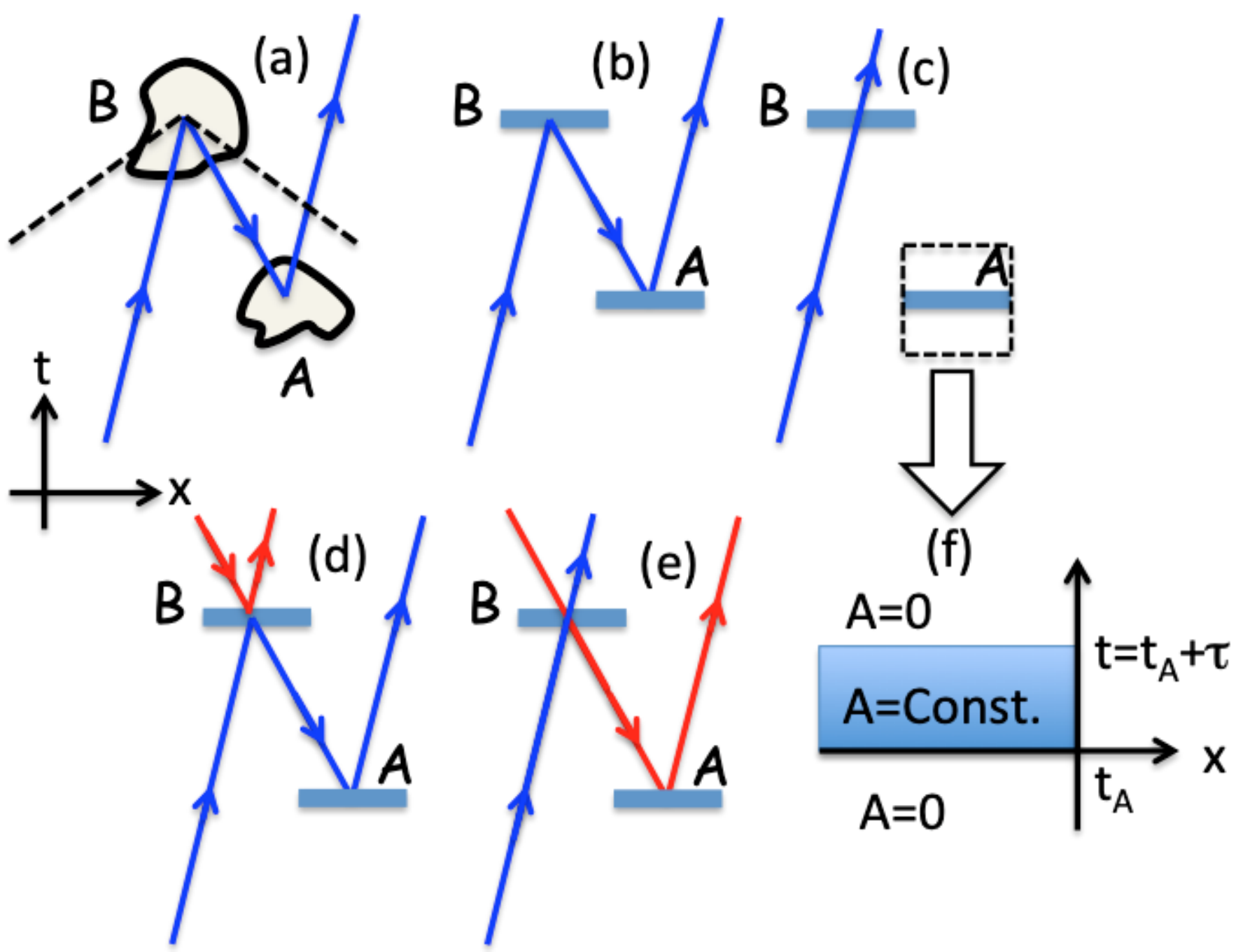}
    \caption{(a) sketch of the `single' electron ziz-zag Feynman's diagram with two disjoints interaction zones $\mathcal{A,B}$.  The dashed lines limits the backward light-cone  from $\mathcal{B}$. (b-c) the two main Feynman's diagrams for the analytical example developed in the text. The two horizontal lines locate the temporal Fabry-Perot cavity sketched in (f): The cavity in $\mathcal{A}$ is characterized by a constant magnetic potential $\mathbf{A}$ in the interval $t\in[t_A,t_A+\tau]$ and vanishing outside this interval. The same is true for the cavity at $\mathcal{B}$ (d-e) Feynman's diagrams involving a electron-positron pair generated in   $\mathcal{A}$ or $\mathcal{B}$ and interfering with the incident electron.   }
    \label{fig:2}
\end{figure}
 \indent In order to avoid paradoxes and problems with causality we stress that the one electron wave function $\psi(x)$ is actually a probability amplitude defined  relatively to the amplitude  $C_v$ of a vacuum field remaining a vacuum under the influence of a potential~\cite{Feynman1949}. This vacuum to vacuum amplitude reads $C_v=\langle\varnothing|\hat{U}(T,0)|\varnothing\rangle$  where $\hat{U}(T,0)$ is the scattering unitary matrix operator (between the initial time $t=0$ and the final time $t=T\rightarrow +\infty$) and $|\varnothing\rangle$ defines the filled Dirac sea. Interestingly $\psi(x)$ can be defined as a weak value~\cite{W1} linked to the two-states formalism~\cite{W2}:\begin{eqnarray}
 \psi(x):=\langle\varnothing;T|\hat{\Psi}(x)\hat{F}^\dagger|\varnothing\rangle/\langle\varnothing;T|\varnothing\rangle
\end{eqnarray} where $\hat{\Psi}(x)$  is a fermionic field operator in the Heisenberg representation~\cite{remark2}, $\hat{F}^\dagger=\int d^3\mathbf{x}\hat{\Psi}^\dagger(\mathbf{x})\psi_0(\mathbf{x})$ is a creation operator for the mode $\psi_0(x)$ and $|\varnothing;T\rangle=U(0,T)|\varnothing\rangle$ is a backward in time evolving quantum state that approaches $|\varnothing\rangle$ at time $t=T$.  Moreover~\cite{Feynman1949,Bjorken1}, relative scattering amplitudes for an electron ending in a positive energy mode $w_n(\mathbf{x})$ are given by $r_n=\int d^3\mathbf{x}w_n^\dagger(\mathbf{x})\psi(\mathbf{x},T)$. The absolute probability amplitudes are  thus given by $r_nC_v$ due to the presence of vacuum. Using this formalism we obtain, in agreement with hole theory, a transition probability $P_n=P_v|r_n|^2$ with $P_v=|C_v|^2$.  Importantly,  the  probabilities of all alternatives must sum to one:
 \begin{eqnarray}
 1=P_v+\sum_n P_n+ \sum P(1 e^- + \textrm{at least 1 pair})\label{norma}
\end{eqnarray}  where we have to include all contributions where electron-positron pairs can be created from vacuum~\cite{Feynman1949}. It is the presence of these pairs interfering with the incident electron wave scattered by the field that preserves causality and prohibits a backward in time signaling.\\
\indent In the following we examines this issue for an idealized system that is treated analytically and will subsequently be used in a QS implementation.  We start with a normalized bispinor plane wave  in a mode volume $V$ \cite{notev}: \begin{eqnarray}
\psi_{E,\mathbf{P}}(t,\mathbf{x})=\sqrt{\frac{E+m}{2EV}}\left(\begin{array}{c}
  \chi\\ \frac{\boldsymbol{\sigma}\cdot\mathbf{p}}{E+m}\chi\end{array}\right)e^{i\mathbf{p}\cdot\mathbf{x}}e^{-iEt}
\end{eqnarray}  with $E=\sqrt{p^2+m^2}$, and $\chi=\left(\begin{array}{c}
  a\\ b\end{array}\right)$ a constant spinor with $\chi^\dagger\chi=1$. This wave with positive energy is impinging on a temporal Fabry-Perot cavity located in region $\mathcal{B}$ at $t=t_\mathcal{B}$ (see Figs.~\ref{fig:2}(b-f)). The cavity at $\mathcal{B}$ (respectively $\mathcal{A}$) is a spacetime domain limited by two hyperplanes at times $t=t_\mathcal{B,A}$ and $t=t_\mathcal{B,A}+\tau$  ($\tau$ is a delay) and where exists a constant magnetic potential $\mathcal{A}$ vanishing outside the domain. As shown in Fig.~\ref{fig:2}(b), the wave can be reflected as a negative energy electron  $re^{-i2Et_\mathcal{B}}\psi_{-E,\mathbf{P}}(t,\mathbf{x})$ propagating backward in time  ($r$ is the Fresnel reflection coefficient for an electron with positive energy  moving forward in time) into the direction of a second identical cavity located at time $t_\mathcal{A}$ where it is again reflected forward in time as $r're^{-i2E(t_\mathcal{B}-t_\mathcal{A})}\psi_{E,\mathbf{P}}(t,\mathbf{x})$ ($r'=-r$ is the reflection coefficient for an electron with negative energy $-E$ moving backward in time). The accumulated phase $e^{-i2E(t_\mathcal{B}-t_\mathcal{A})}$ is reminiscent of Feynman's zigzag diagram~\cite{Feynman1949}.  As shown in Fig.~\ref{fig:2}(c), there is a second channel where the electron is directly transmitted by the 
$\mathcal{B}-$cavity as a positive energy wave $t\psi_{-E,\mathbf{P}}(t,\mathbf{x})$.  The Fabry-Perot cavities are characterized by four coefficient $r,r',t,t'$ that are calculated by usual means (see technical details in \cite{SupM}). In particular we have for the reflectivity $R=|r|^2=|r'|^2=R'$ and transmissivity $T=|t|^2=|t'|^2=T'$ with  $R+1=T$.\\
\indent The probability for the electron to be reflected (Fig.~\ref{fig:2}(b)) is given by $P_{reflec.}=P_v|rr'|^2=P_vR^2$ and the one to be transmited (Fig.~\ref{fig:2}(c)) is $P_{trans.}=P_v|t|^2=P_v(1+R)$ where $P_v$ is the vacuum to vacuum probability. In order to have normalization of the probability (as in Eq.~\ref{norma}) we must include two other Feynman diagrams shown in Figs.~\ref{fig:2}(d,e) associated with an additional pair creation in either $\mathcal{A}$ or $\mathcal{B}$. The two diagrams interfere and a relative minus sign is introduced to satisfy Pauli's exclusion principle.  The  additional probability term is thus $P_{e^-+ 1 pair}=P_v|rr'r-r'tt'|^2=P_vR|rr'-tt'|^2=P_vR$~\cite{SupM}. The total probability is the sum 
\begin{eqnarray}
P_{tot.}=P_v(R^2+1+2R)=P_v(1+R)^2.
\end{eqnarray} 
In order to prove that $P_{tot.}=1$ we have to evaluate $P_v$. This is done by considering the 3 alternatives where in absence of the incoming electron (i) nothing happens with probablity $P_v$, (ii) a pair is created at $\mathcal{B}$ with probability $P_v|r'|^2=P_vR$, or (iii) a pair is created at $\mathcal{A}$ with probability $P_v|t'r|^2=P_vRT$. The sum of (i-iii) gives 
\begin{eqnarray}
1=P_v(1+R+RT)=P_v(1+R)^2 \label{normproba}
\end{eqnarray}  and therefore we have the normalization $P_{tot.}=1$ as required. Note that Feynman's scattering  method has been previously applied to solve the famous Klein paradox~\cite{Klein} or the related Schwinger pair production problem~\cite{pairprob}.\\  
\begin{figure}[h]
    \centering
    \includegraphics[width=0.8\linewidth]{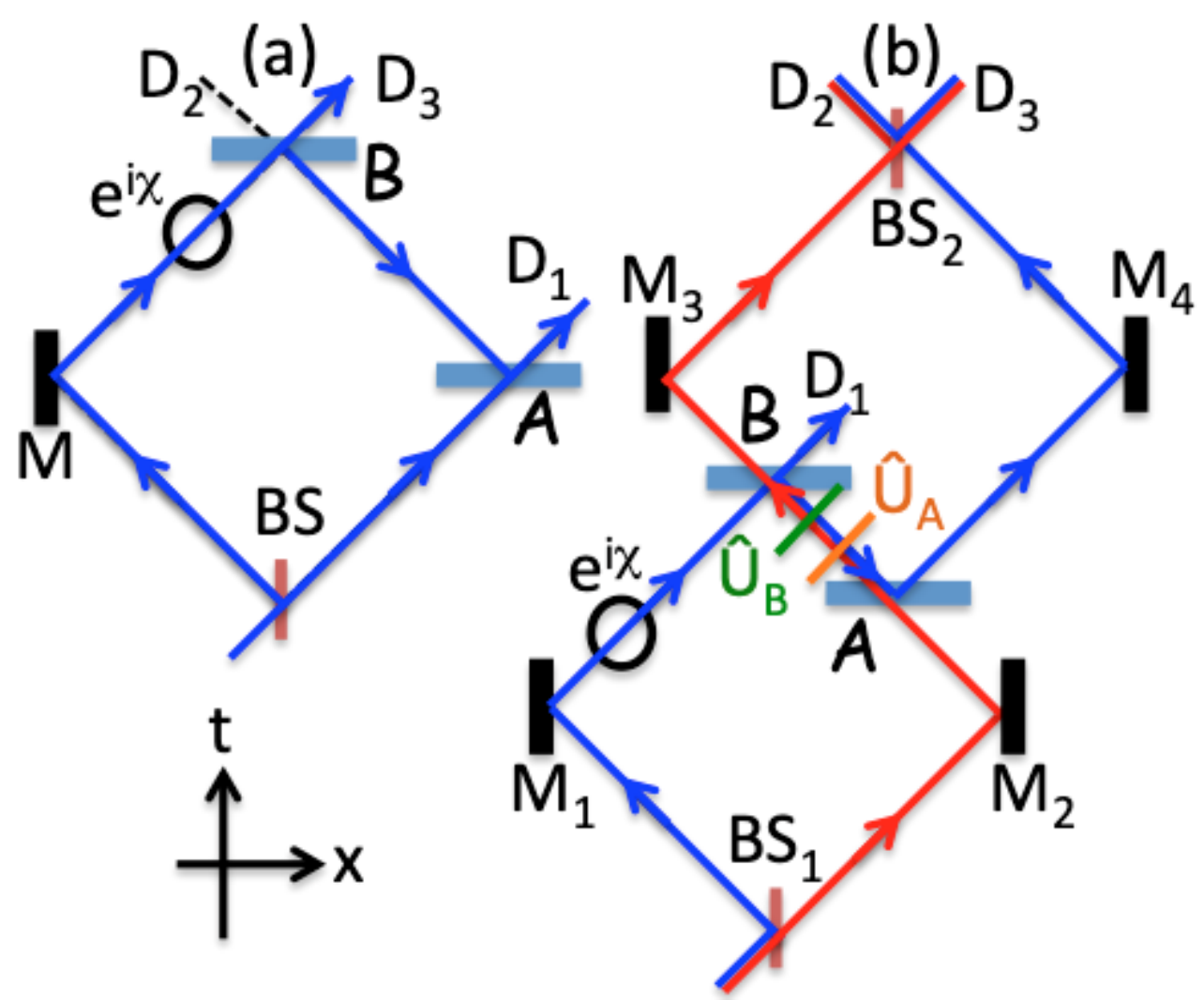}
    \caption{(a) sketches a retrocausal electron interferometer involving the two temporal cavities $\mathcal{A},\mathcal{B}$ ($BS$ is a beam splitter, $M$ a mirror, $e^{i\chi}$ a phase shifter, and $D_i$ detectors). (b) retrocausal QS involving two cavities $\mathcal{A},\mathcal{B}$. Unitaries $\hat{U}_A,\hat{U}_B$ acting on the spinor states are located in the beams  ($BS_i,M_i,D_i$ are respectively beamsplitters, mirrors, and detectors). The two paths in red and blue correspond to the two causal order of Fig.~\ref{fig:coordinate}(a).}
    \label{fig:3}
\end{figure}
\indent In order to see the impact of the previous analysis on causality we consider the interferometer of Fig.~\ref{fig:3}(a) where a single electron ($E>0$) impinging on a beam splitter $BS$ is following one of the three paths (i) $BS\prec\mathcal{A}\prec D_1$, (ii) $BS\prec M\prec\mathcal{B}\prec\mathcal{A}\prec D_1$, or (iii) $BS\prec M\prec\mathcal{B}\prec D_3$ with $M$ a mirror and $D_i$ detectors. Paths (i) and (ii) interfere and by introducing a phase shift $e^{i\chi}$ in (ii) we can modulate the intensity at $D_1$ through a retrocausal link involving the cavities $\mathcal{B,A}$. The probability of path (iii) to detect an electron at $D_3$ and no particle at $D_{1,2}$ ($D_2$ is a positron detector) is given  by $P(D_3,\varnothing D_{1,2})=P_vT/2$ where $P_v$ is as before given by Eq.~\ref{normproba}. Paths (i) and (ii) lead to the modulated probability of finding an electron in gate $D_1$ and no particle in gates $D_{2,3}$, i.e.:  
\begin{eqnarray}
P(D_1,\varnothing D_{2,3})=\frac{P_v}{2}|t-ie^{i(\chi-2E(t_\mathcal{B}-t_\mathcal{A}))}r^2|^2\nonumber\\
=\frac{P_v}{2}(T+R^2)(1+\mathcal{V}\sin{\theta})\label{b}
\end{eqnarray}  with visibility $\mathcal{V}=\frac{2R\sqrt{T}}{T+R^2}\leq 1$ and phase $\theta=\chi-2E(t_\mathcal{B}-t_\mathcal{A})+2\arg{[r]}-\arg{[t]}$ as we show in~\cite{SupM}.\\
\indent This interference illustrates the retrocausal mechanism associated with the presence of zigzag Feynman diagrams in Figs.~\ref{fig:2} and \ref{fig:3}. In particular the phase can always be tuned to $\theta_\pm=\pm\pi/2$ and if $\mathcal{V}=1$ (implying the `golden ratio'-reflectivity $R_0=\frac{1+\sqrt{5}}{2}\simeq 1.62$~\cite{SupM}) we have $P_{\theta_-}(D_1,\varnothing D_{2,3})=0$. Therefore, the presence of the phase shifter at time $t~\simeq t_{\mathcal{B}}$ strongly retrocausally influences the dynamics of the particle at time $t_{\mathcal{A}}<t_{\mathcal{B}}$.\\
\indent Remarkably, this feature cannot be used to send a signal back to the past. Indeed, we neglected  Feynman's diagrams where a pair of particle-antiparticle is created either at $\mathcal{A}$ or $\mathcal{B}$. Similar to those shown in  Figs.~\ref{fig:2}(d,e) we obtain three additional interfering Feynman's graphs that yield the probability for finding an electron at $D_1$ and a pair of particle  at $D_{2,3}$~\cite{SupM} 
\begin{eqnarray}
P_{\theta_-}(D_1,D_{2},D_3)=\frac{P_vR}{2}[2+R-2\sqrt{T}\sin{\theta}].\label{c}
\end{eqnarray} 
The $\theta-$oscillating term in Eq.~\ref{c} exactly compensates the one in Eq.~\ref{b} so that the full probability to detect one electron at $D_1$ (i.e., obtained by summing Eqs.~\ref{b},\ref{c}) is independent of $\theta$. An observer at $D_1$ not knowing what happens at $D_{2,3}$ will thus detect a constant probability. This is a form of non-signalling theorem (similar to the one involved in discussing Bell's theorem~\cite{Bell,Nosig}) protecting quantum mechanics from backward in time information transfer. This is imposed by local commutativity and microcausality in QED where local measurements of quantities  $\hat{M}_1$ and $\hat{N}_2$ made at points $1$ and $2$ must commute, i.e., $[\hat{M}_1,\hat{N}_2]_-=0$, if they can be connected by a space-like hypersurface passing through $1$ and $2$. Here, this is so if detectors $D_1$ and $D_{2,3}$ are located on  such a space-like hypersurface.\\
\indent Moreover, like for Bell's theorem, correlations and postselections are key and can be used to retrodict a backward influence.  For this purpose, consider the following game where a fair quantum coin allows us to select randomly  between the case $\theta_-$ and $\theta_+$ (with $\mathcal{V}=1$).  A single electron is sent through the interferometer and the agent having access to the outcomes at $D_i$ must guess the result of the coin tossing and therefore $\theta$. Calling $a=\pm \pi/2$ the guess we introduce an average Gain \begin{eqnarray}
\langle G\rangle=\sum_{a,\theta}\delta_{a,\theta}P(a,\theta)
\end{eqnarray} where the joint probablity $P(a,\theta)$ depends on information available to the agent. Classically, i.e., without the retrocausal channel, the agent could not use the data at $D_i$ to infer $\theta$ and therefore she would have to toss a coin to guess  yielding the classical  bound \begin{eqnarray}
\langle G\rangle_{clas.}\leq\frac{1}{4}\sum_{a,\theta}\delta_{a,\theta}=\frac{1}{2}.\label{causal}
\end{eqnarray}
   Moreover, as we show in \cite{SupM}, in our retrocausal quantum model  we obtain 
\begin{eqnarray}
\langle G\rangle_{quant.}=\frac{3}{4}-\frac{R_0}{4(1+R_0)^2}\simeq 0.69> \frac{1}{2}.
\end{eqnarray} Note that all data and correlations have been used and  our causal inequality Eq.~\ref{causal} is clearly violated. Furthermore, by postselecting only on those events where a single detection at $D_1$ occurs and none at $D_{2,3}$ we can increase the gain to reach the maximal value $\langle G\rangle_{|D_1,\varnothing D_{2,3}}:=\sum_{a,\theta}\delta_{a,\theta}P(\theta|D_1,\varnothing D_{2,3})=1$ (see \cite{SupM} for details).\\
\indent \textit{Discussion-} Going back to the issue of ICO, the  previous analysis allows us to develop genuine versions of QS involving only two times regions $\mathcal{A}$, $\mathcal{B}$ as required in Fig.~\ref{fig:coordinate}(a). On the example of Fig.~\ref{fig:3}(b) a single electron moving through an interferometer and reaching a detector $D_2$ or $D_3$ follows either a `normal' path $BS_1\prec M_2\prec \mathcal{A}\prec \mathcal{B}\prec M_3\prec D_{2,3}$  or the zig-zag one $BS_1\prec M_1\prec \mathcal{B}\prec\mathcal{A}\prec M_4\prec D_{2,3}$. Unitaries $\hat{U}_A,\hat{U}_B$ acting on the bi-spinor quantum state of the electron $\psi_0$ are inserted in  the paths joining  $\mathcal{A,B}$ and the interference  at $D_{2}$ and $D_3$ leads to ICO. The probability  for detecting a single electron at $D_2$  and nothing at $D_{3,1}$ is:   
\begin{eqnarray}
P(D_2,\varnothing D_{3,1})=\frac{P'_v}{4}\sum_s\lVert t^2C_s^{\mathcal{A}\prec\mathcal{B}}-r^2e^{i\xi}C_s^{\mathcal{B}\prec\mathcal{A}}\rVert ^2\label{ICO}
\end{eqnarray}  with $\xi=\chi-2E(t_\mathcal{B}-t_\mathcal{A})$ and here $P'_v(1+R)^4=1$ because we need two counterpropagating electron modes.  The amplitudes $C_s^{\mathcal{A}\prec\mathcal{B}}$, $C_s^{\mathcal{B}\prec\mathcal{A}}$  ($s=\pm\frac{1}{2}$ are spin states) depend on the order of the unitaries $\hat{U}_A,\hat{U}_B$ applied on $\psi_0$ as we show  in \cite{SupM}.\\ 
\indent Remarkably, unlike previous proposals this QS exemplifies ICO but with definite time order. This shines some new lights concerning the distinction between time and causal order in relativistic quantum mechanics. Moreover, our analysis  also impacts the discussion of causal inequalities that are usually non violated with QS~\cite{Oreshkov2012,Branciard2016,Purves2021} at least if we assume no-backward causation~\cite{Yelena2019}. But this is precisely the condition that we relax here and as a consequence causal inequalities can be violated even with a definite time order.\\
\indent Consider for example the standard bipartite `guess your neighbor's input' (GYNI) game~\cite{Oreshkov2012,Branciard2016,Mafalda} where Alice and Bob with their respective (input) quantum bits with results $x,y$ (and probabilities $P(x)=P(y)=\frac{1}{2}$) try to guess each  other's input. The gain in this game with outputs $a,b$ is thus~\cite{Branciard2016}  
\begin{eqnarray}
\langle G\rangle=\frac{1}{4}\sum_{a,b,x,y}\delta_{a,y}\delta_{b,x}P(a,b|x,y),\label{causalb}
\end{eqnarray}  where the conditional probability $P(a,b|x,y)$  depends on causal rules. Suppose that Alice's lab $\mathcal{A}$ is in the past light cone of Bob's lab $\mathcal{B}$ we know that Alice can communicate  with certainty her input $x$ to Bob (i.e., $b=x$) but Bob cannot and Alice must guess $y$ by using a random number so  that $P(a,b|x,y)\leq \delta_{b,x}\frac{1}{2}$ leading to the causal bound  $\langle G\rangle_{caus.}\leq\frac{1}{2}$.  However, if Bob uses a single electron with a well controlled bispinor state $\psi_0$ he can send information backward  in time to Alice by acting unitarily on $\psi_0$ and using a zigzag channel like the one of Fig.~\ref{fig:2}(b)   (Alice would have to analyse the bispinor state). As we saw the probability of this single electron channel requiring postselection is $P_{reflec.}=P_vR^2$ and we now have $P(a,b|x,y)= \delta_{b,x}\delta_{a,y}P_{reflec.}$  yielding
\begin{eqnarray}
\langle G\rangle_{retrocaus.}=P_{reflec.}=\frac{R^2}{(1+R)^2}.\label{causalc}
\end{eqnarray}Furthermore, if $R\geq R_1:= 1+\sqrt{2}$ we have $\langle G\rangle_{retrocaus.}\geq \frac{1}{2}$ violating the causal inequality. By studying numerically~\cite{SupM} $R$ as a function of the external potential $\mathbf{A}$ and imput particle energy $E$ we can reach the value $\langle G\rangle_{retrocaus.}\simeq 0.98$. We stress that in this work values for $R\gtrsim R_0,R_1$ require very strong external field beyond the Schwinger limit~\cite{Schwinger1951} $10^{18}$ V/m during times of the order of the Compton period $\sim m^{-1}$ \cite{SupM}. In the present knowledge these could be attainable in extreme physical phenomena like supernova or Blackholes.\\
\indent Once again, we emphasize that all our results rely on correlations in order to avoid problems with violation of non-signalling and microcausality. This should be compared with other retrodictive/retrocausal results in quantum mechanics such as the quantum delayed choice quantum eraser~\cite{Scully} or entanglement swapping~\cite{Ma} that however, unlike our work, not use the `retrocausal' Feynman propagator $S_F(x-x')$.\\
\indent Finally, we mention that by combining temporal cavites like $\mathcal{A,B}$ with other linear devices we can implement closed time-like curves (CTCs) that have been strongly  debated in the recent years~\cite{Deutsch,Greenberger,Lloyd,Ringbauer} outside the field of general relativity from which it emerged~\cite{Thorne1}. Like in experimental simulations based on quantum teleportation~\cite{Lloyd} our CTCS based on  electron loops in QED require correlations and postselections in order to solve the `grandfather paradox' in a consistent way without violating nosignalling (a short analysis is provided in \cite{SupM}). In that sense the `chronology protection conjecture' advocated by Hawking~\cite{Hawking} is here replaced by a non-signalling condition protecting our macroscopic local world from quantum CTCs. 
 

\newpage
\pagebreak
\widetext
\begin{center}
\textbf{\large Supplemental Materials: Indefinite causal order with fixed temporal order for electrons and positrons}
\end{center}

\section*{Summary}
\indent This suplementary file details the calculations contained in the main article:  In  Sections~\ref{Sec1}, \ref{Sec2} and \ref{Sec3} the properties of the temporal Fabry-Perot involving negative energy electrons moving  backward in time are derived. In  Section~\ref{Sec4} the principle of Feynman scattering formalism \cite{Feynman1949} are applied to the calculation of detection probabilities for exmamples  discussed in the main article. In Section \ref{Sec5} we discuss a retrocausal game violating some inequalites and in Section \ref{Sec6} we derive the working conditions for a quantum switch involving zigzag retrocausal paths. We end with a discussion in Section \ref{Sec7} of closed time-like curves (CTC) in connections with retrocausal trajectories. 
\section{Appendix 1: The single $t=0$ interface}\label{Sec1}
\begin{figure}[h]
    \centering
    \includegraphics[width=6cm]{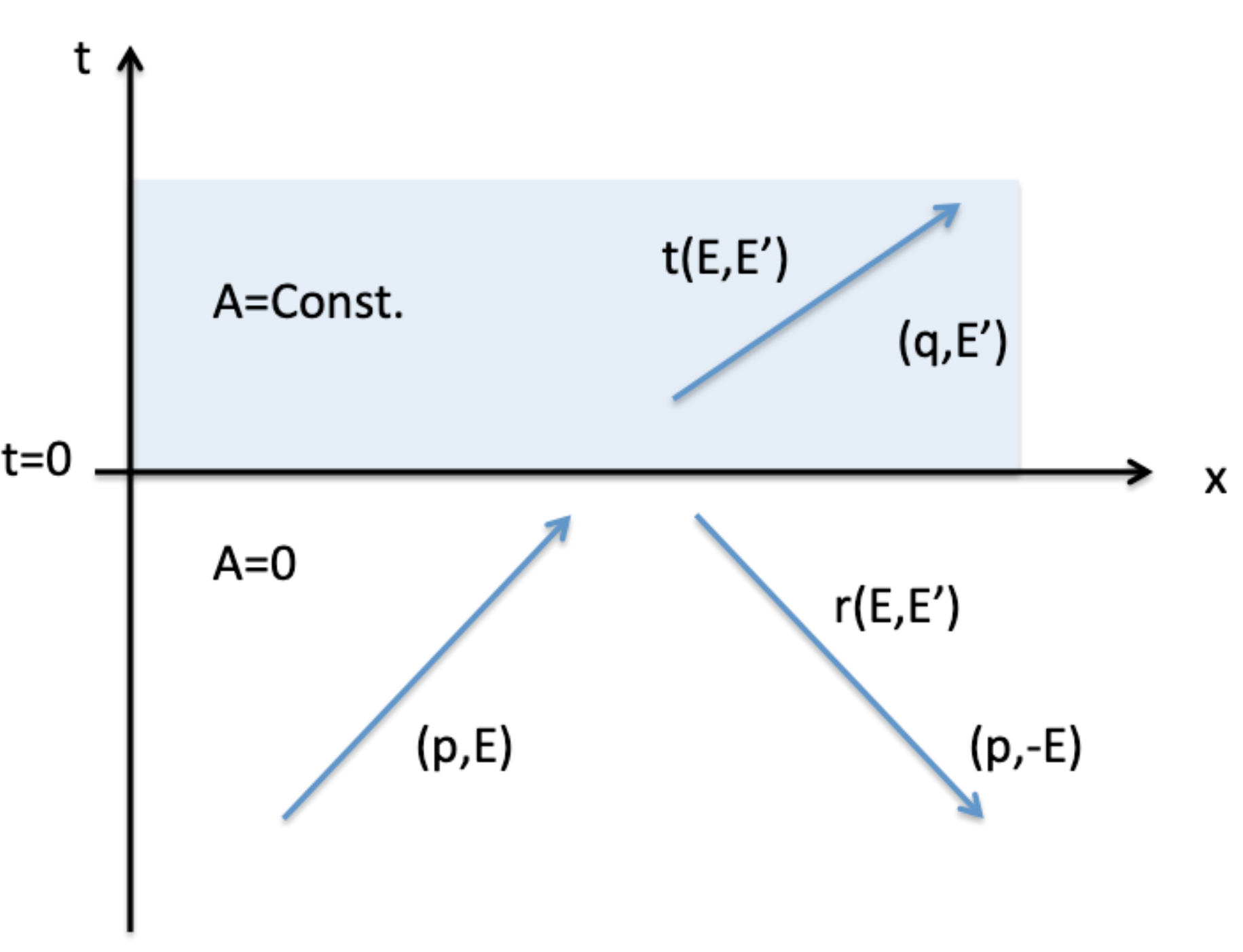}
    \caption{ Scattering of a single relativistic electron at an interface spearating a region where $\mathbf{A}=0$ from a region where $\mathbf{A}=const.$. The negative energy mode is associated with a particle moving backward in time.  }
    \label{fig:1}
\end{figure}
We start from Dirac's equation in a constant magnetic potential field and we define a configuration where a fermion be spinor associated with an electron with positive energy $E$ and momentum $\mathbf{p}$ (i.e., a plane wave) is reflected at time $t=0$ by an electromagnetic field discontinuity (see Fig.\ref{fig:1}).  The electron is reflected as a positron identified with an electron with the same momentum $\mathbf{p}$ (due to conservation of momentum at the interface $t=0$) and the negative energy $-E$ moving backward in time. The transmitted wave is an electron   with energy $+E'$ and momentum  $\mathbf{p}$ (again due to conservation of momentum at the interface). More precisely:  If $t<0$ we have
\begin{eqnarray}
\Psi_{<}(\mathbf{x},t)=\sqrt{\frac{E+m}{2EV}}\left(\begin{array}{c}
  \chi\\ \frac{\boldsymbol{\sigma}\cdot\mathbf{p}}{E+m}\chi
  \end{array}\right)e^{i\mathbf{p}\cdot\mathbf{x}}e^{-iEt}+r(E,E')\sqrt{\frac{E-m}{2EV}}\left(\begin{array}{c}
  \chi\\ -\frac{\boldsymbol{\sigma}\cdot\mathbf{p}}{E-m}\chi
  \end{array}\right)e^{i\mathbf{p}\cdot\mathbf{x}}e^{iEt}\label{eq1}
\end{eqnarray} (with $\chi^\dagger\chi=1$) and if $t>0$ we have
\begin{eqnarray}
\Psi_{>}(\mathbf{x},t)=t(E,E')\sqrt{\frac{E'+m}{2E'V}}\left(\begin{array}{c}
  \chi\\ \frac{\boldsymbol{\sigma}\cdot\mathbf{q}}{E'+m}\chi
  \end{array}\right)e^{i\mathbf{p}\cdot\mathbf{x}}e^{-iE't},\label{eq2}
\end{eqnarray}  where $\mathbf{q}=\mathbf{p}-e\mathbf{A}$, $E=\sqrt{\mathbf{p}^2+m^2}$, $E'=\sqrt{\mathbf{q}^2+m^2}$ and $\mathbf{A}$ is a magnetic vector potential defined as constant for $t>0$ and vanishing for $t<0$. $r(E,E')$ (respectively $t(E,E')$) defines the Fresnel reflection (transmission) coefficient; $V$ is the volume of a large  box introduced for normalization of the waves. Physically we have a discontinuous electric field $\mathbf{E}(t)=-\partial_t\mathbf{A}(t)=-\delta(t)\mathbf{A}$, and a vanishing magnetic field $\mathbf{B}=0$.  Maxwell's equations impose a discontinuous electric current $\mathbf{J}(t)=-\partial_t\mathbf{E}(t)=\frac{d\delta(t)}{dt}\mathbf{A}$ located on the hyperplane $t=0$.\\ 
We have the normalization 
$\int d^3\mathbf{x}\Psi_{<}^\dagger(\mathbf{x},t)\Psi_{<}(\mathbf{x},t)=1+|r(E,E')|^2$, $\int d^3\mathbf{x}\Psi_{>}^\dagger(\mathbf{x},t)\Psi_{>}(\mathbf{x},t)=|t(E,E')|^2$. The continuity at $t=0$ 
\begin{eqnarray}
\Psi_{<}(\mathbf{x},t=0)=\Psi_{>}(\mathbf{x},t=0)\label{eq3}
\end{eqnarray}
imposes $1+|r(E,E')|^2=|t(E,E')|^2$
i.e.,   with $T(E,E'):=|t(E,E')|^2$ and $R(E,E'):=|R(E,E')|^2$ the probability conservation 
\begin{eqnarray}
1+R(E,E')=T(E,E'),\label{eqtra}
\end{eqnarray} where $R(E,E')$ is the reflectivity and $T(E,E')$ the transmissivity.  Observe that Eq.~\ref{eqtra} is different from the usual rule `$R+T=1$' used in optics where we consider interfaces in space but not in time.
The Fresnel coefficients $r(E,E')$ and $t(E,E')$ are obtained from Eqs.~\ref{eq1}, \ref{eq2}, \ref{eq3}:
\begin{eqnarray}
r(E,E')=\sqrt{\frac{E+m}{E-m}}\frac{E-m-\frac{pq}{E'+m}}{E+m+\frac{pq}{E'+m}}\nonumber\\
t(E,E')=1+r(E,E')=\sqrt{\frac{E+m}{E'+m}}\frac{2E}{E+m+\frac{pq}{E'+m}}\label{fresnel}
\end{eqnarray}
with $p=\sqrt{E^2-m^2}$ and $q=\sqrt{E'^2-m^2}$.\\
\begin{figure}[h]
    \centering
    \includegraphics[width=6cm]{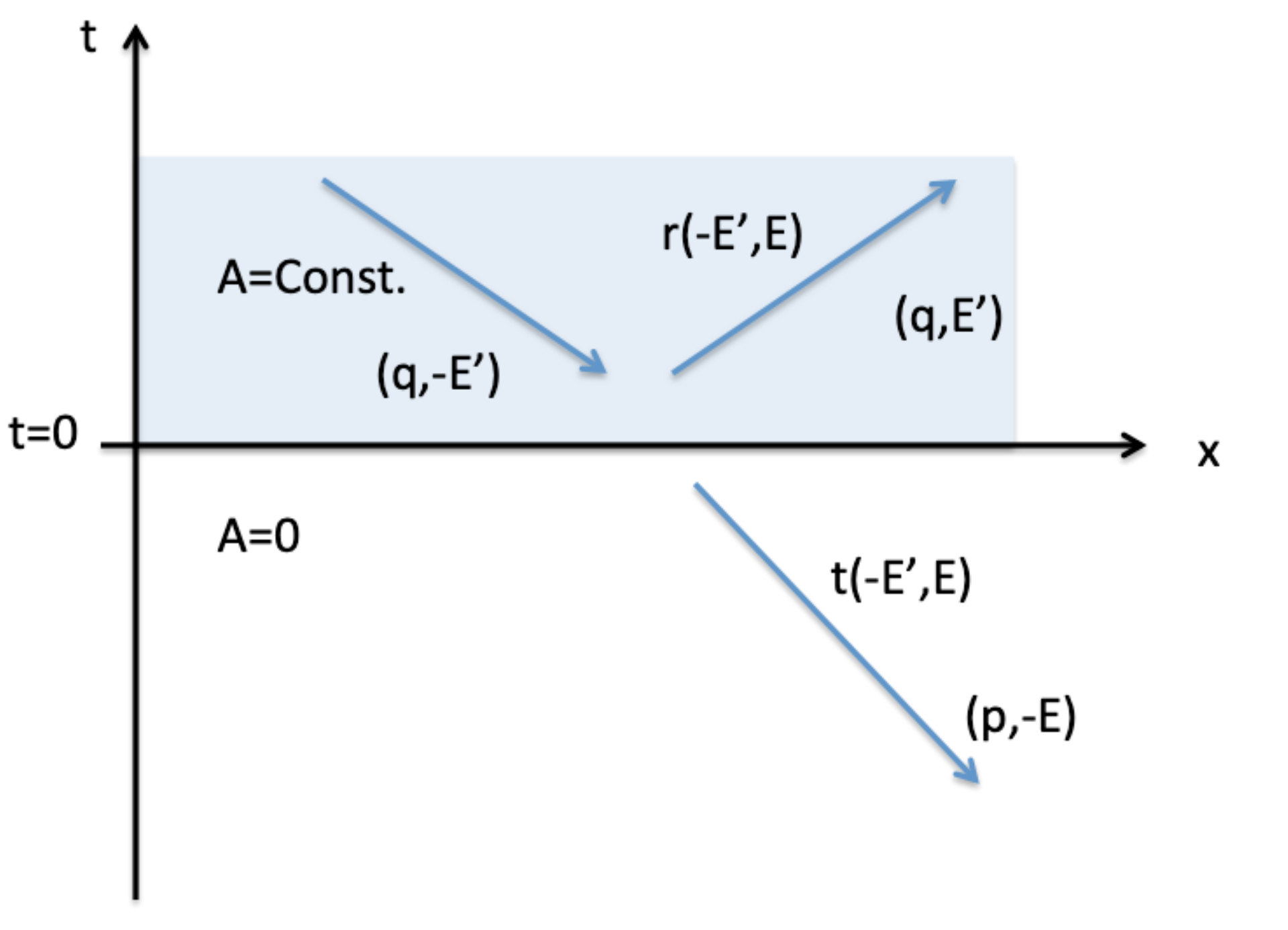}
    \caption{Same as in Fig.\ref{fig:1} but with the incident electron coming from the region $\mathbf{A}=const.$ }
    \label{fig:2}
\end{figure}
\indent Similarly we can define a configuration  sketched in Fig.~\ref{fig:2} where we have an incoming positron (an electron with negative energy $-E'$ going backward in time) inciend from $t>0$ that is reflected at the interface $t=0$ where it becomes an electron with positive energy $+E'$ propagating forward in time. The transmitted wave (i.e., $t<0$) evolves as a positron (i.e. an electron with energy $-E$ going backward in time).  The magnetic potential is defined as before.  The fermion fields are now defined as: 
\begin{eqnarray}
\Psi_{>}(\mathbf{x},t)=\sqrt{\frac{E'-m}{2E'V}}\left(\begin{array}{c}
  \chi\\ -\frac{\boldsymbol{\sigma}\cdot\mathbf{q}}{E'-m}\chi
  \end{array}\right)e^{i\mathbf{p}\cdot\mathbf{x}}e^{+iE't}+r(-E',E)\sqrt{\frac{E'+m}{2E'V}}\left(\begin{array}{c}
  \chi\\ \frac{\boldsymbol{\sigma}\cdot\mathbf{q}}{E'+m}\chi
  \end{array}\right)e^{i\mathbf{p}\cdot\mathbf{x}}e^{-iE't},
\end{eqnarray} for $t>0$ and 
\begin{eqnarray}
\Psi_{<}(\mathbf{x},t)=t(-E',E)\sqrt{\frac{E-m}{2EV}}\left(\begin{array}{c}
  \chi\\ -\frac{\boldsymbol{\sigma}\cdot\mathbf{p}}{E-m}\chi
  \end{array}\right)e^{i\mathbf{p}\cdot\mathbf{x}}e^{iEt}
\end{eqnarray} for $t<0$.   Here the Fresnel coefficients are obtained from Eq.~\ref{fresnel} after the transformation $E\rightarrow-E'$, $E'\rightarrow-E$ and $p\leftrightarrow q$.\\
\begin{figure}[h]
    \centering
    \includegraphics[width=9cm]{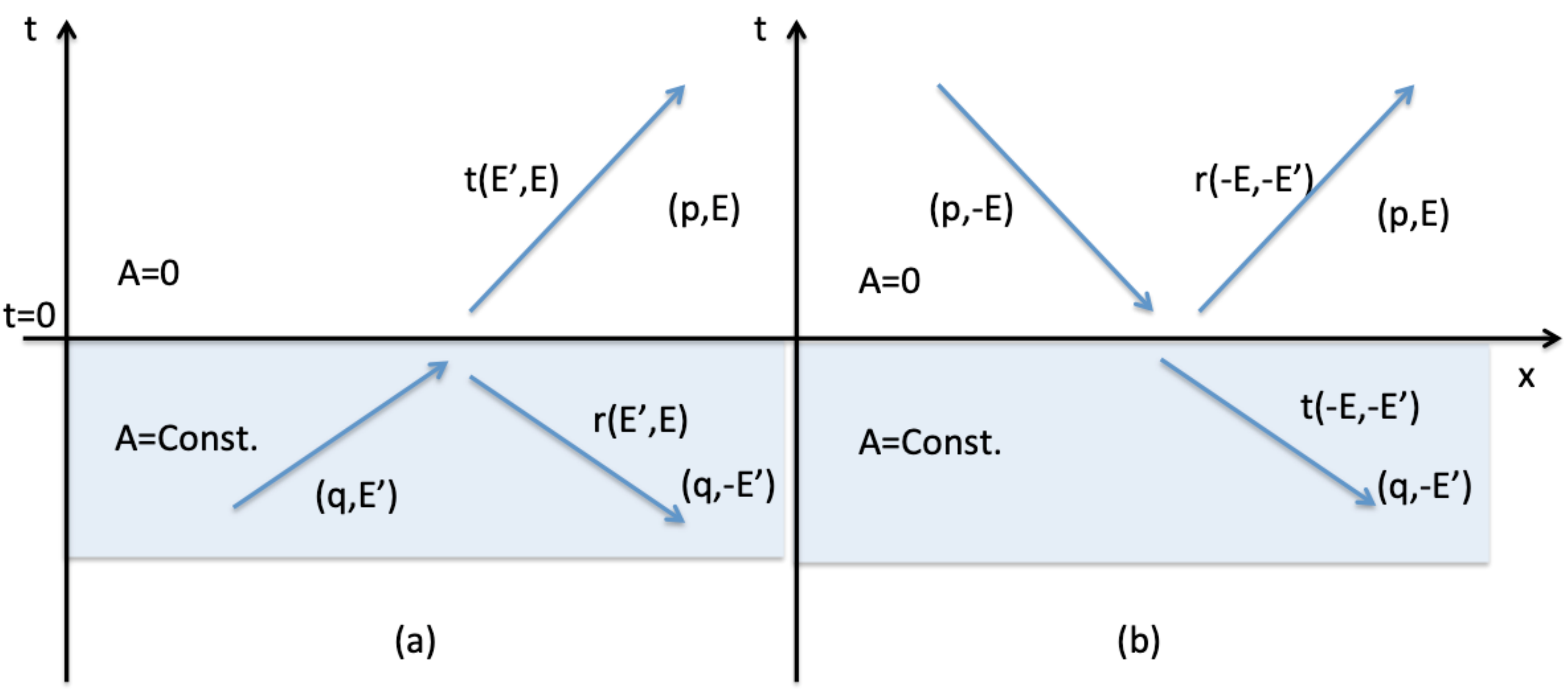}
    \caption{ Other configurations where the regions $\mathbf{A}=0$ and $\mathbf{A}=const.$ are swapped compared to Figs.\ref{fig:1},\ref{fig:2}. }
    \label{fig:3}
\end{figure}
\indent We also define two configurations where the magnetic potential is defined by the same constant vector $\mathbf{A}$ but now for $t<0$.  For $t>0$ the potential is vanishing (see Fig.~\ref{fig:3}). In the configuration shown in Fig.~\ref{fig:3}(a) an incident electron with positive energy $+E'$ is reflected at the interface at time $t=0$ and changed into a reflected negative energy particle with the same momentum $\mathbf{p}$. The transmitted wave for $t>0$ has a positive energy $+E$.  The fermion fields are defined as:  
\begin{eqnarray}
\Psi_{<}(\mathbf{x},t)=\sqrt{\frac{E'+m}{2E'V}}\left(\begin{array}{c}
  \chi\\ \frac{\boldsymbol{\sigma}\cdot\mathbf{q}}{E'+m}\chi
  \end{array}\right)e^{i\mathbf{p}\cdot\mathbf{x}}e^{-iE't}+r(E',E)\sqrt{\frac{E'-m}{2E'V}}\left(\begin{array}{c}
  \chi\\ -\frac{\boldsymbol{\sigma}\cdot\mathbf{q}}{E'-m}\chi
  \end{array}\right)e^{i\mathbf{p}\cdot\mathbf{x}}e^{iE't}
\end{eqnarray} and if $t>0$ we have
\begin{eqnarray}
\Psi_{>}(\mathbf{x},t)=t(E',E)\sqrt{\frac{E+m}{2EV}}\left(\begin{array}{c}
  \chi\\ \frac{\boldsymbol{\sigma}\cdot\mathbf{p}}{E+m}\chi
  \end{array}\right)e^{i\mathbf{p}\cdot\mathbf{x}}e^{-iEt}.
\end{eqnarray}
 This times the Fresnel coefficients $r(E',E)$ and $t(E',E)$ are related to Eq.~\ref{fresnel} by the transformation $E\rightarrow E'$,$E'\rightarrow E$, and $p\leftrightarrow q$. The last configuration shown in Fig.~\ref{fig:3}(b) corresponds to an electron with negative energy $-E$ incident from the region $t>0$ and reflected as a positive electron with energy $+E$ and the same momentum $\mathbf{p}$. The transmitted wave is a negative energy electron (i.e. with energy $-E'$).  We have for the fermion fields:
 \begin{eqnarray}
\Psi_{>}(\mathbf{x},t)=\sqrt{\frac{E-m}{2EV}}\left(\begin{array}{c}
  \chi\\ -\frac{\boldsymbol{\sigma}\cdot\mathbf{p}}{E-m}\chi
  \end{array}\right)e^{i\mathbf{p}\cdot\mathbf{x}}e^{+iEt}+r(-E,-E')\sqrt{\frac{E+m}{2EV}}\left(\begin{array}{c}
  \chi\\ \frac{\boldsymbol{\sigma}\cdot\mathbf{p}}{E+m}\chi
  \end{array}\right)e^{i\mathbf{p}\cdot\mathbf{x}}e^{-iEt},
\end{eqnarray} for $t>0$ and 
\begin{eqnarray}
\Psi_{<}(\mathbf{x},t)=t(-E,-'E)\sqrt{\frac{E'-m}{2E'V}}\left(\begin{array}{c}
  \chi\\ -\frac{\boldsymbol{\sigma}\cdot\mathbf{q}}{E'-m}\chi
  \end{array}\right)e^{i\mathbf{p}\cdot\mathbf{x}}e^{iE't}
\end{eqnarray} for $t<0$. The Fresnel coefficient are now given by Eq.~\ref{fresnel} and the transformation $E\rightarrow -E$,$E'\rightarrow -E'$,  ($p,q$ are unchanged). We mention a useful property (see Appendix \ref{Sec3}):
\begin{eqnarray}
r(-E,-E')=-r(E,E').\label{interesting}
\end{eqnarray}  \\
\indent We stress that the electron/positron waves propagating into the region where the constant potential is $\mathbf{A}$ are actually waves propagating in vacuum as it can be checked by introducing a gauge transformation.  More precisely,   if we consider Dirac's equation $\gamma^\mu(i\partial_\mu-A_\mu(x))\psi(x)=m\psi(x)$ in presence of a four vector potential $A(x)$ we have the $U(1)$ local gauge invariance:
$\psi(x)=\psi'(x)e^{ie\chi(x)}$, $A'_\mu(x)=A_\mu(x)+\partial_\mu\chi(x)$ leading to $\gamma^\mu(i\partial_\mu-A'_\mu(x))\psi'(x)=m\psi'(x)$. Therefore, if we have a fermion wave $\psi(x)=e^{-ipx}\phi$ (where $\phi$ is a constant bispinor and $p$ a four vector momentum) in a constant four vector potential $A$ we have \begin{eqnarray}
\gamma^\mu(p_\mu-A_\mu)\psi(x)=m\psi(x).\label{gauge}
\end{eqnarray}  Using the gauge function $\chi(x)=-A_\mu x^\mu$  to cancel  we get $\psi'(x)=e^{-iqx}\phi$, $A'=0$  where $q=p-eA$ is the new four vector momentum of the particle and the wave equation reads 
\begin{eqnarray}
\gamma^\mu q_\mu\psi'(x)=m\psi'(x).
\end{eqnarray} that is equivalent to Eq.~\ref{gauge}. 
\section{Appendix 2: A temporal Fabry-Perot cavity for electron/positron waves }\label{Sec2}
\subsection{Theory}
\indent In the next step we consider a magnetic potential $\mathbf{A}(t)$ such that $\mathbf{A}(t)=0$ for $t<0$ and $t>\tau>0$ and $\mathbf{A}(t)=\mathbf{A}=const.$ for $0<t< \tau$ (see Fig.~\ref{fig:4}). Using the results obtained in the Appendix \ref{Sec1} we can define a Fabry-Perot cavity in time for electron / positron waves, i.e., involving electron waves with positive energys $E,E'$ moving forward in time and negative energy $-E,-E'$ moving backward in time.    
\begin{figure}[h]
    \centering
    \includegraphics[width=8cm]{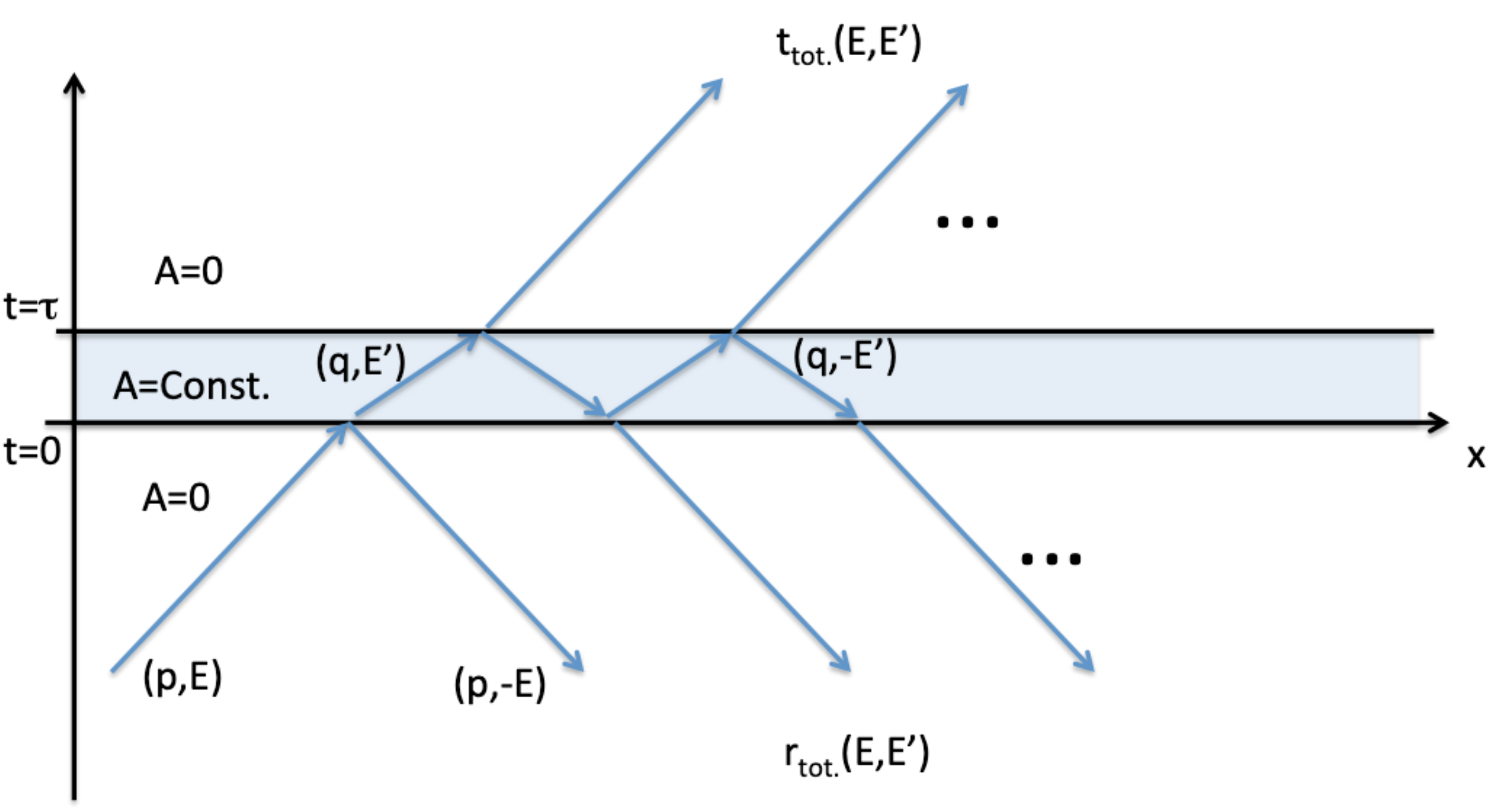}
    \caption{ A temporal Fabry-Perot cavity acting as a beam splitter using a  temporal region where $\mathbf{A}=const.$ sandwiched between two semi-infinite regions $\mathbf{A}=0$. The incident electron is coming from the region $t=-\infty$. }
    \label{fig:4}
\end{figure}
Consider first the case of a electron wave such that for $t<0$ we have  
\begin{eqnarray}
\Psi_{<}(\mathbf{x},t)=\sqrt{\frac{E+m}{2EV}}\left(\begin{array}{c}
  \chi\\ \frac{\boldsymbol{\sigma}\cdot\mathbf{p}}{E+m}\chi
  \end{array}\right)e^{i\mathbf{p}\cdot\mathbf{x}}e^{-iEt}+r_{tot.}(E,E')\sqrt{\frac{E-m}{2EV}}\left(\begin{array}{c}
  \chi\\ -\frac{\boldsymbol{\sigma}\cdot\mathbf{p}}{E-m}\chi
  \end{array}\right)e^{i\mathbf{p}\cdot\mathbf{x}}e^{iEt}\label{eq1tot}
\end{eqnarray} and if $t>\tau$ we have
\begin{eqnarray}
\Psi_{>}(\mathbf{x},t)=t_{tot.}(E,E')\sqrt{\frac{E+m}{2EV}}\left(\begin{array}{c}
  \chi\\ \frac{\boldsymbol{\sigma}\cdot\mathbf{p}}{E+m}\chi
  \end{array}\right)e^{i\mathbf{p}\cdot\mathbf{x}}e^{-iEt}.\label{eq2tot}
\end{eqnarray}
For the time interval $0<t<\tau$ we have a superposition of plane waves 
\begin{eqnarray}
\Psi_{cavity}(\mathbf{x},t)=\alpha(E,E')\sqrt{\frac{E'+m}{2E'V}}\left(\begin{array}{c}
  \chi\\ \frac{\boldsymbol{\sigma}\cdot\mathbf{q}}{E'+m}\chi
  \end{array}\right)e^{i\mathbf{p}\cdot\mathbf{x}}e^{-iE't}+\beta(E,E')\sqrt{\frac{E'-m}{2E'V}}\left(\begin{array}{c}
  \chi\\ -\frac{\boldsymbol{\sigma}\cdot\mathbf{q}}{E'-m}\chi
  \end{array}\right)e^{i\mathbf{p}\cdot\mathbf{x}}e^{iE't}\label{eq3tot}
\end{eqnarray} but in the following we are not interested in the coefficients $\alpha(E,E'), \beta(E,E')$.
 The total Fresnel's reflection and transmission coefficients for the time-cavity can be calucated in the usual way by introducing a phase delay $\Delta=-E'\tau$ accumulated by the waves during their propagation between the two  time interfaces. We obtain by iteration and using the results of the  Appendix \ref{Sec1}: 
 \begin{eqnarray}
 r_{tot.}(E,E')=r(E,E')+\frac{t(E,E')r(E',E)t(-E',-E)e^{i2\Delta}}{1-r(E',E)r(-E',-E)e^{i2\Delta}}\nonumber\\
 t_{tot.}(E,E')=\frac{t(E,E')t(E',E)e^{i\Delta}}{1-r(E',E)r(-E',-E)e^{i2\Delta}}.\label{fresneltot1}
\end{eqnarray} 
\begin{figure}[h]
    \centering
    \includegraphics[width=8cm]{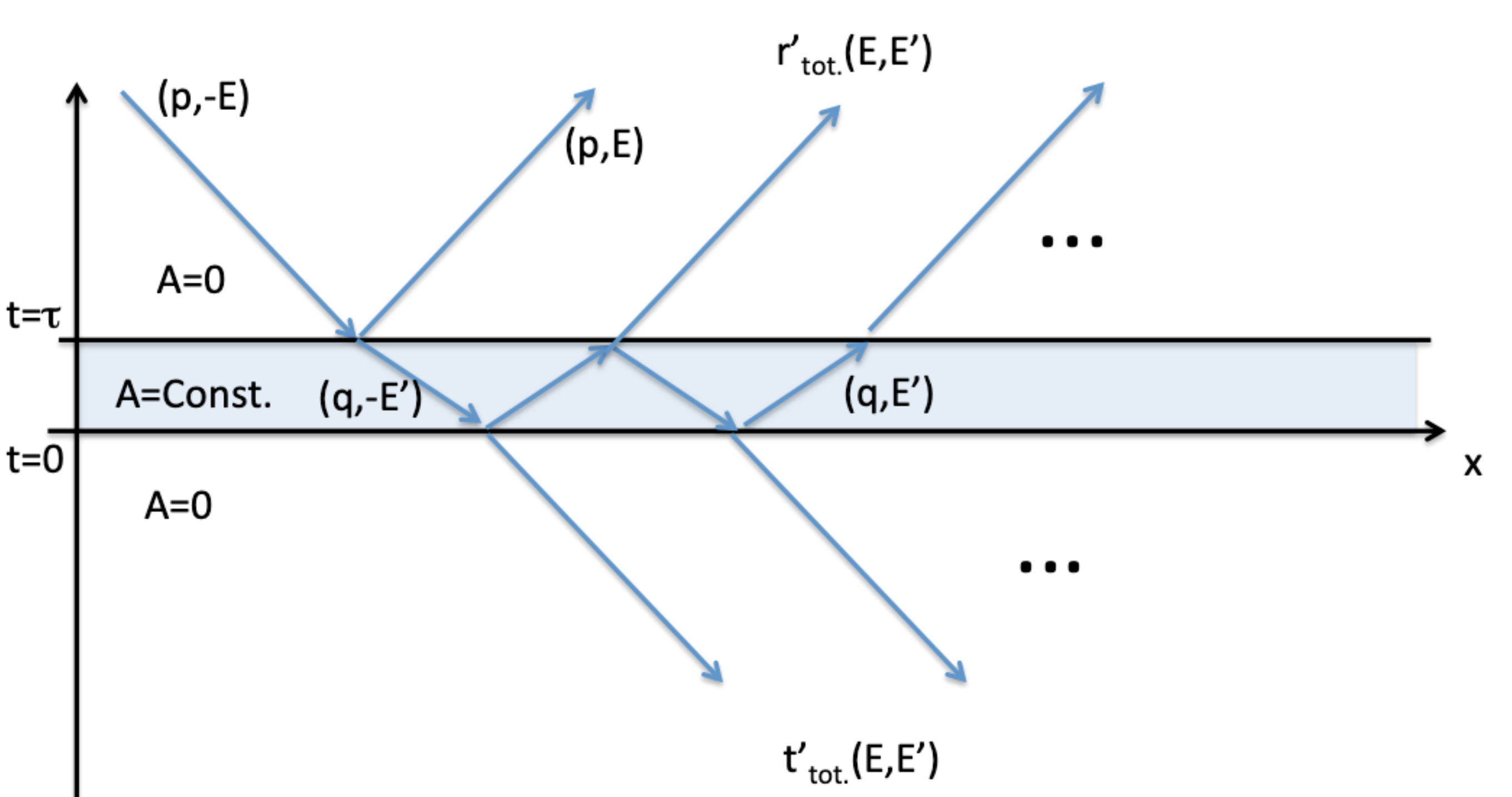}
    \caption{ Same as in Fig.~\ref{fig:4} but with the incident electron coming from the region $t=+\infty$. }
    \label{fig:5}
\end{figure}
We can also consider the configuration where an incident electron with negative energy $-E$ propagates from the region $t>\tau$ in direction of the time cavity. A part will be reflected and transformed into a positive energy electron propagating forward in time and a part will be transmitted in the domain $t<0$ with a negative energy (see Fig.~\ref{fig:5}).   The full wave reads 
\begin{eqnarray}
\Psi_{>}(\mathbf{x},t)=\sqrt{\frac{E-m}{2EV}}\left(\begin{array}{c}
  \chi\\ -\frac{\boldsymbol{\sigma}\cdot\mathbf{p}}{E-m}\chi
  \end{array}\right)e^{i\mathbf{p}\cdot\mathbf{x}}e^{iEt}+r'_{tot.}(E,E')\sqrt{\frac{E+m}{2EV}}\left(\begin{array}{c}
  \chi\\ \frac{\boldsymbol{\sigma}\cdot\mathbf{p}}{E+m}\chi
  \end{array}\right)e^{i\mathbf{p}\cdot\mathbf{x}}e^{-iEt}\label{eq1totb}
\end{eqnarray} for $t>\tau$ and \begin{eqnarray}
\Psi_{<}(\mathbf{x},t)=t'_{tot.}(E,E')\sqrt{\frac{E-m}{2EV}}\left(\begin{array}{c}
  \chi\\ -\frac{\boldsymbol{\sigma}\cdot\mathbf{p}}{E-m}\chi
  \end{array}\right)e^{i\mathbf{p}\cdot\mathbf{x}}e^{iEt}.\label{eq2tot}
\end{eqnarray} for $t<0$. This leads to the Fresnel's coefficients 
\begin{eqnarray}
 r'_{tot.}(E,E')=r(-E,-E')+\frac{t(-E,-E')r(-E',-E)t(E',E)e^{i2\Delta}}{1-r(E',E)r(-E',-E)e^{i2\Delta}}\nonumber\\
 t'_{tot.}(E,E')=\frac{t(E,E')t(E',E)e^{i\Delta}}{1-r(E',E)r(-E',-E)e^{i2\Delta}}.\label{fresneltot2}
\end{eqnarray}  
\subsection{Numerical properties of the reflectivity and transmission}
\begin{figure}[h]
    \centering
    \includegraphics[width=8cm]{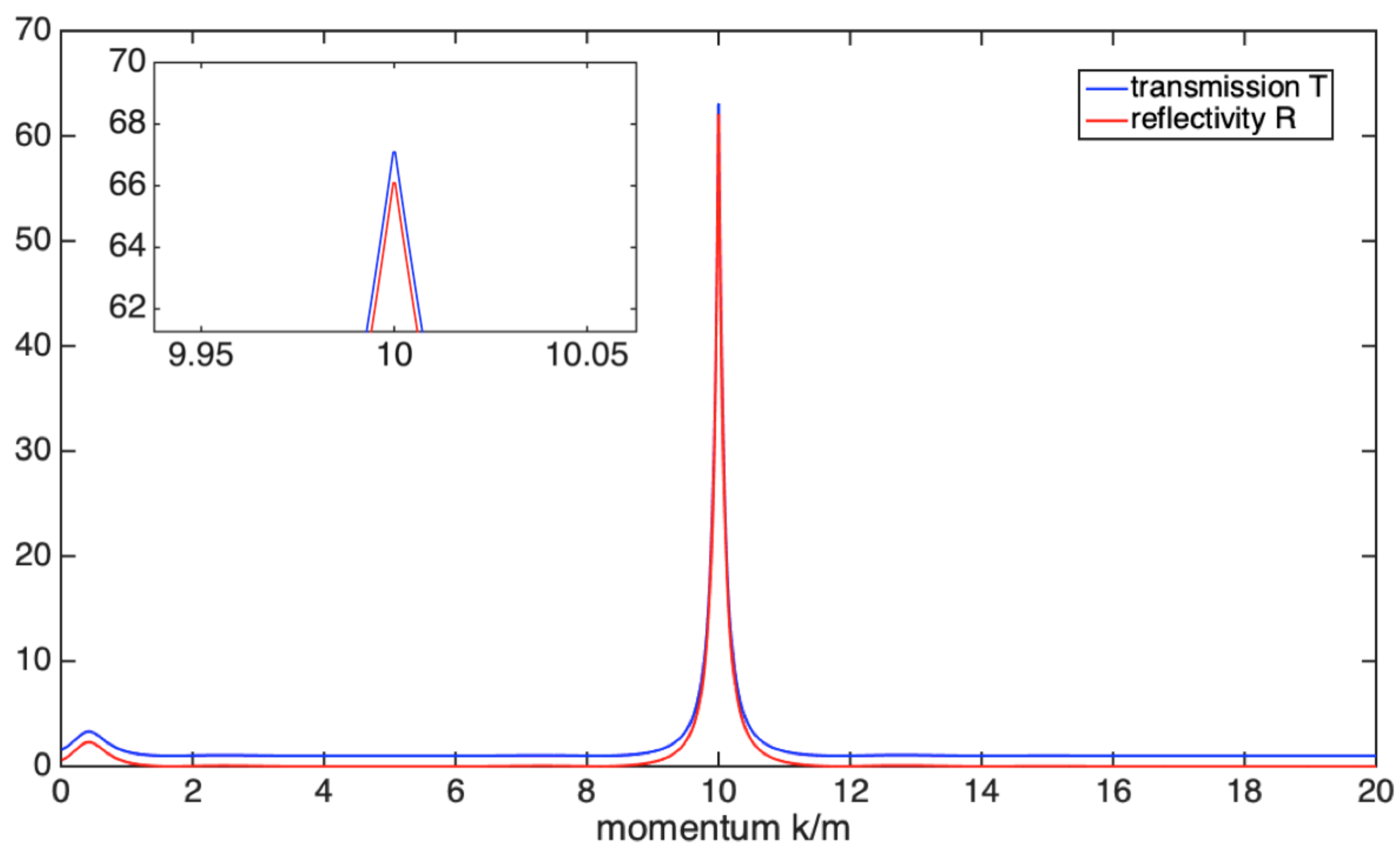}
    \caption{ Properties of the coefficients $R_{tot.}(E,E')=|r_{tot.}(E,E')|^2$ and $T_{tot.}(E,E')=|t_{tot.}(E,E')|^2(=1+R_{tot.}(E,E'))$ as a function of $|\mathbf{k}|/m$ for a particular relevant case with  $m\tau=1.5$ and $|e\mathbf{A}|/m=10$. The inset shows a zoom.}
    \label{fig:5b}
\end{figure}
\begin{figure}[h]
    \centering
    \includegraphics[width=9cm]{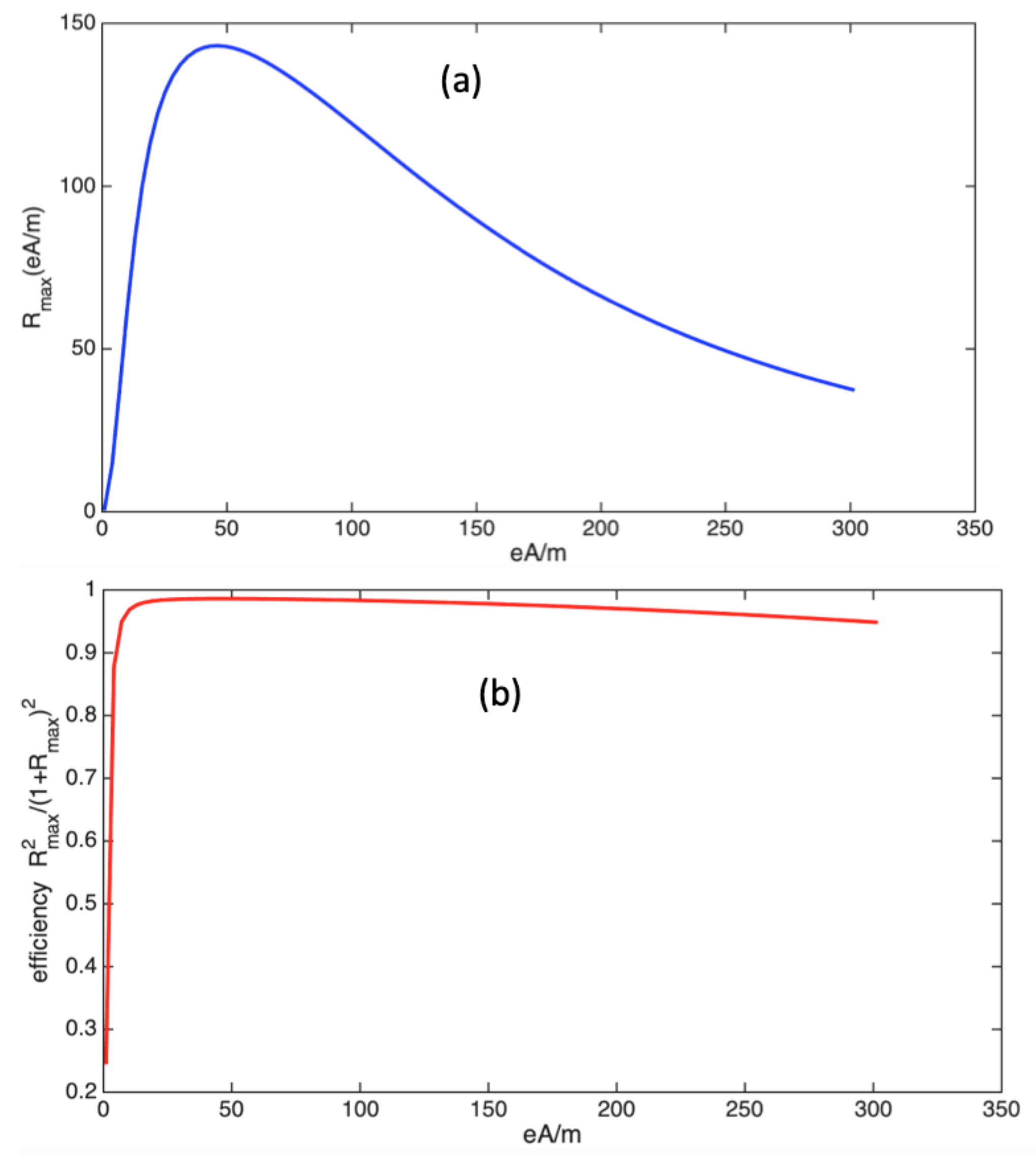}
    \caption{ (a) the maximum reflectivity $R_{max}:=R_{tot.}(E,E')=|r_{tot.}(E,E')|^2$ (obtained for the momentum $|\mathbf{k}|\simeq |e\mathbf{A}|$) as a function of $e|\mathbf{A}|/m=10$ for $m\tau=1.5$. (b) efficiency $R^2/(1+R)^2$ computed for $R=R_{max}$ as a function of $e|\mathbf{A}|/m=10$ for $m\tau=1.5$. }
    \label{fig:5c}
\end{figure}
In the present work we are interested in the properties $R_{tot.}(E,E')=|r_{tot.}(E,E')|^2$ and $T_{tot.}(E,E')=|t_{tot.}(E,E')|^2(=1+R_{tot.}(E,E'))$ for regimes where $R_{tot.}(E,E')> 1$. We observed that this possible only if $e|\mathbf{A}|/m>1$.   In Fig.~\ref{fig:5b} we show an illustration for the case $m\tau= 1.5$ and $e|\mathbf{A}|/m=10$.  In this regime the maximum of reflectivity $R_{max}$  is observed for $|\mathbf{k}|\simeq |e\mathbf{A}|$.   We studied the evolution of this maximum of reflectivity $R_{max}$ as function of $e|\mathbf{A}|/m>$ and as shown in Fig.~\ref{fig:5c}(a) the reflectivity can reach very high value up to $R_{max}\simeq 143.13$ for a field $e|\mathbf{A}|/m\simeq 46.45$. This issue is particularly in the context of the  `guess your neighbor's input' (GYNI) game~\cite{Oreshkov2012,Branciard2016,Mafalda} discussed in the main article.  Indeed,  the gain depends on the efficiency \begin{eqnarray}
\eta=\frac{R_{tot.}(E,E')^2}{(1+R_{tot.}(E,E'))^2}.
\end{eqnarray}In the regime considered here we see that for $R_{max}:=R_{tot.}(E,E')$ the efficiency can reach very high value. This is shown in Fig.~\ref{fig:5c}(b) where $\eta$ can reach the value $\eta\simeq 0.9862$ for a field $e|\mathbf{A}|/m\simeq 46.45$. We stress that the condition $\eta\geq \frac{1}{2}$ discussed in the main article implies  $R\geq 1+\sqrt{2}$.\\
\indent In this context it is interesting to observe that physically our temporal Fabry-Perot Cavity is an idealization corresponding to infinite slopes at times $t=0$ and $t=\tau$.  Infinite slopes actually means that the electric field $\mathbf{E}(t)=-\partial_t\mathbf{A}(t)$  equals $\mathbf{E}(t)=-\mathbf{A}\delta(t)$ near the first spacelike interface  $t=0$ and  $\mathbf{E}(t)=+\mathbf{A}\delta(t-\tau)$ near the second spacelike interface where $\mathbf{A}$ is the constant magnetic potential value in the cavity.   This is of course an idealization  but we must have $|\mathbf{E}|\simeq |\mathbf{A}|/\varepsilon$ near $t=0$  and $t=\tau$ where $\varepsilon\ll \tau$.  In other words,  we have $|e\mathbf{E}|\gg |e\mathbf{A}|/\tau$ and consequently if $m\tau\sim 1$ we have the constraint 
\begin{eqnarray}
|\mathbf{E}|\gg \frac{m^2}{|e|}
\end{eqnarray} which is the limit  obtained  by Schwinger for pair productions in a constant electric field~\cite{Schwinger1951}.  This limit for electron/positron pair:  $\frac{m^2}{|e|}\simeq 10^{+18} V.m^{-1}$ is an extremely intense field currently unatainable with our technology but supposed to be possible during extreme astrophysical/cosmological events.  
\section{Appendix 3: Parity-time inversion symmetry }\label{Sec3}
\indent We remind that the solutions of Dirac's equation evolve under a parity transformation $\mathbf{x}'=-\mathbf{x}$, $t'=t$ as
\begin{eqnarray}
\psi'(\mathbf{x}',t'):=\psi_P(-\mathbf{x},t)=\gamma^0\psi(\mathbf{x},t)
\end{eqnarray}
with \begin{eqnarray}
\gamma^0=\left(\begin{array}{cc}
  \mathbb{1}&0\\
  0&-\mathbb{1}
  \end{array}\right)=\left(\begin{array}{cccc}
  1&0&0&0\\
  0&-1&0&0\\
  0&0&-1&0\\
  0&0&0&1
  \end{array}\right).
\end{eqnarray}
Similarly  the time inversion symmetry $\mathbf{x}'=+\mathbf{x}$, $t'=-t$  leads to the transformation
\begin{eqnarray}
\psi'(\mathbf{x}',t'):=\psi_T(\mathbf{x},-t)=i\gamma^1\gamma^3\psi^\ast(\mathbf{x},t)
\end{eqnarray}
with \begin{eqnarray}
i\gamma^1\gamma3=\left(\begin{array}{cc}
  -\sigma_y&0\\
  0&-\sigma_y
  \end{array}\right)=i\left(\begin{array}{cccc}
  0&1&0&0\\
  -1&0&0&0\\
  0&0&0&1\\
  0&0&-1&0
  \end{array}\right).\end{eqnarray}
  Applying these two transformations sequentially on a plane wave solution 
  \begin{eqnarray}
\psi(\mathbf{x},t)=\sqrt{\frac{E+m}{2EV}}\left(\begin{array}{c}
  \chi\\ \frac{\boldsymbol{\sigma}\cdot\mathbf{p}}{E+m}\chi
  \end{array}\right)e^{i\mathbf{p}\cdot\mathbf{x}}e^{-iEt}\label{spinor}
\end{eqnarray}  with $\chi:=\left(\begin{array}{c}
  a\\ b
  \end{array}\right)$, leads to
\begin{eqnarray}
\psi'(\mathbf{x}',t'):=\psi_{TP}(-\mathbf{x},-t)=i\gamma^2\gamma^3\gamma^0\psi^\ast(\mathbf{x},t)=
\sqrt{\frac{E+m}{2EV}}\left(\begin{array}{c}
  -\sigma_y\chi^\ast\\ -\frac{\boldsymbol{\sigma}\cdot\mathbf{p}}{E+m}\sigma_y\chi^\ast
  \end{array}\right)e^{i\mathbf{p}\cdot\mathbf{x}'}e^{-iEt'}.
\end{eqnarray}This shows that a plane wave with a momentum four vector $p:=(E,\mathbf{P})$ (with $E$ positive or negative) is transformed into the same plane wave at the new point $x'=-x:=(-t,-\mathbf{x})$ and with a new spinor state $\chi'=-\sigma_y\chi^\ast$. Clearly if we apply this TP transformation to a plane state  
\begin{eqnarray}
\psi(\mathbf{x},t)=\sqrt{\frac{E+m}{2EV}}\left(\begin{array}{c}
  \phi\\ \frac{\boldsymbol{\sigma}\cdot\mathbf{p}}{E+m}\phi
  \end{array}\right)e^{i\mathbf{p}\cdot\mathbf{x}}e^{-iEt}
\end{eqnarray}  with $\phi:=+\sigma_y\chi^\ast$ we obtain after the TP transformation the state  
\begin{eqnarray}
\psi'(\mathbf{x}',t')=
\sqrt{\frac{E+m}{2EV}}\left(\begin{array}{c}
  \chi\\ \frac{\boldsymbol{\sigma}\cdot\mathbf{p}}{E+m}\chi
  \end{array}\right)e^{i\mathbf{p}\cdot\mathbf{x}'}e^{-iEt'}.
\end{eqnarray}  with the same spinor $\chi$ as in Eq.~\ref{spinor}.\\
\indent This  TP(=PT) transformation can be applied to the problem of the two previous sections for finding some symmetries of the Fresnel coefficient. For this it is important to remember that under TP a electromagnetic potential four vector $A(x)$  transforms as $A'(x')=A(x)=A(-x').$
We consider in details the application to the problem of the time-cavity of Appendix \ref{Sec2}. For this purpose we start with the configuration of Fig.~\ref{fig:4} leading to the Fresnel coefficients $r_{tot.}(E,E'),t_{tot.}(E,E')$ of Eq.~\ref{fresneltot1}. 
\begin{figure}[h]
    \centering
    \includegraphics[width=10cm]{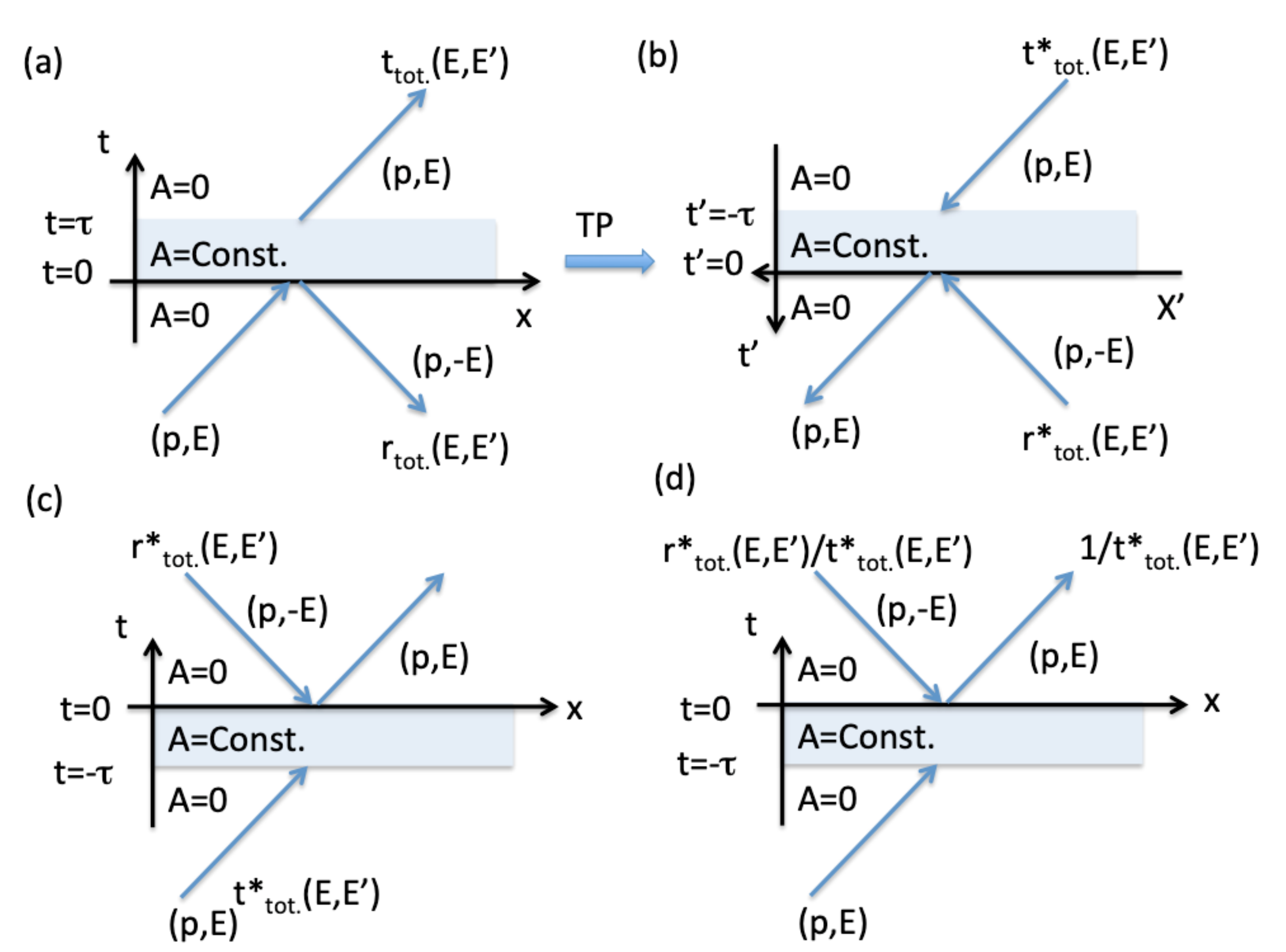}
    \caption{ TP transformation applied to a single electron scattered by a temporal Fabry Perot cavity (see text). }
    \label{fig:6}
\end{figure}
As shown in Fig.~\ref{fig:6}(a) and Fig.~\ref{fig:6}(b) the TP transformation let the time-cavity invariant but change the fresnel coefficients in the different beams (note that in Fig.~\ref{fig:6}(b) the arrows for the electron/positron convention are inversed with respect to Fig.~\ref{fig:6}(a) due to the time inversion).  This (passive) transformation is equivalent to the active transformation shown in Fig.~\ref{fig:6}(c) where we now replaced variables $t,\mathbf{x}$ by $t',\mathbf{x}'$ (the time-cavity is the same as in Fig.~\ref{fig:6}(a) up to a time translation). Finally if we divide all the amplitudes in the different beams by $t^\ast_{tot.}(E,E')$ we get the result of Fig.~\ref{fig:6}(d) where a negative energy electron coming from $t>0$ and with amplitude $r^\ast_{tot.}(E,E')/t^\ast_{tot.}(E,E')$ propagates backward in time  and interferes with a positive energy electron (coming from $t<-\tau$) with a unit amplitude. The result of the interference leads  to the  reflected positive energy electron propagating forward in time in the domain $t>0$ with an amplitude $1/t^\ast_{tot.}(E,E')$.\\
\indent This result should be compared with the results for  $r_{tot.}(E,E'),t_{tot.}(E,E')$ of Eq.~\ref{fresneltot1} and $r'_{tot.}(E,E'),t'_{tot.}(E,E')$ of Eq.~\ref{fresneltot2}. In order to be consistent we easily obtain the two conditions:
\begin{eqnarray}
|t_{tot.}(E,E')|^2+r'_{tot.}(E,E')r^\ast_{tot.}(E,E')=1\nonumber\\
r_{tot.}(E,E')t^\ast_{tot.}(E,E')+r^\ast_{tot.}(E,E')t'_{tot.}(E,E')=0.\label{truc1}
\end{eqnarray}  
Moreover, we must have the conservation rule $|t_{tot.}(E,E')|^2=1+|r_{tot.}(E,E')|^2$ and therefore from Eq.~\ref{truc1} we have:
\begin{eqnarray}
r'_{tot.}(E,E')=-r_{tot.}(E,E').\label{truc2}
\end{eqnarray} Similarly,  $|t'_{tot.}(E,E')|^2=1+|r'_{tot.}(E,E')|^2=1+|r_{tot.}(E,E')|^2=|t_{tot.}(E,E')|^2$ and thus  we have
\begin{eqnarray}
|t'_{tot.}(E,E')|=|t_{tot.}(E,E')|.\label{truc3}
\end{eqnarray}
Finally, using Eqs.~\ref{truc1},\ref{truc2} and writing $t_{tot.}(E,E')=|t_{tot.}(E,E')|e^{i\varphi_t}$, $r_{tot.}(E,E')=|r_{tot.}(E,E')|e^{i\varphi_r}$, $t'_{tot.}(E,E')=|t'_{tot.}(E,E')|e^{i\varphi_{t'}}$ we obtain 
\begin{eqnarray}
2\varphi_r-\varphi_t-\varphi_{t'}=\pm\pi.\label{truc4}
\end{eqnarray}
\indent This is actually not the end of the properties we can deduce from the PT transformation. Indeed, applying the same procedure to the two interfaces at $t=0$ and $t=\tau$ we can obtain constraints on the various Fresnel coefficients $r(E,E'), t(E,E')$ deduced in Appendix \ref{Sec1}.
After lengthly calculations that are not reproduced here we can indeed deduce:
\begin{eqnarray}
t(E,E')t(E',E)=t(-E,-E')t(-E',-E).
\end{eqnarray} 
Inserting this relation in Eqs.~\ref{fresneltot1},\ref{fresneltot2} and using Eq.~\ref{interesting} yields
\begin{eqnarray}
t'_{tot.}(E,E')=t_{tot.}(E,E'),\nonumber\\
r'_{tot.}(E,E')=-r_{tot.}(E,E'),
\end{eqnarray} that is in agreement with Eqs.~\ref{truc2},\ref{truc3} if $\varphi_t=\varphi_{t'}$. This implies from  Eq.~\ref{truc4}
   \begin{eqnarray}
\varphi_r-\varphi_t=\pm\pi/2.\label{truc4}
\end{eqnarray}  
\section{Appendix 4: Computing probabilities  }\label{Sec4}
\subsection{The principle~\cite{Feynman1949}}
\begin{figure}[h]
    \centering
    \includegraphics[width=8.5cm]{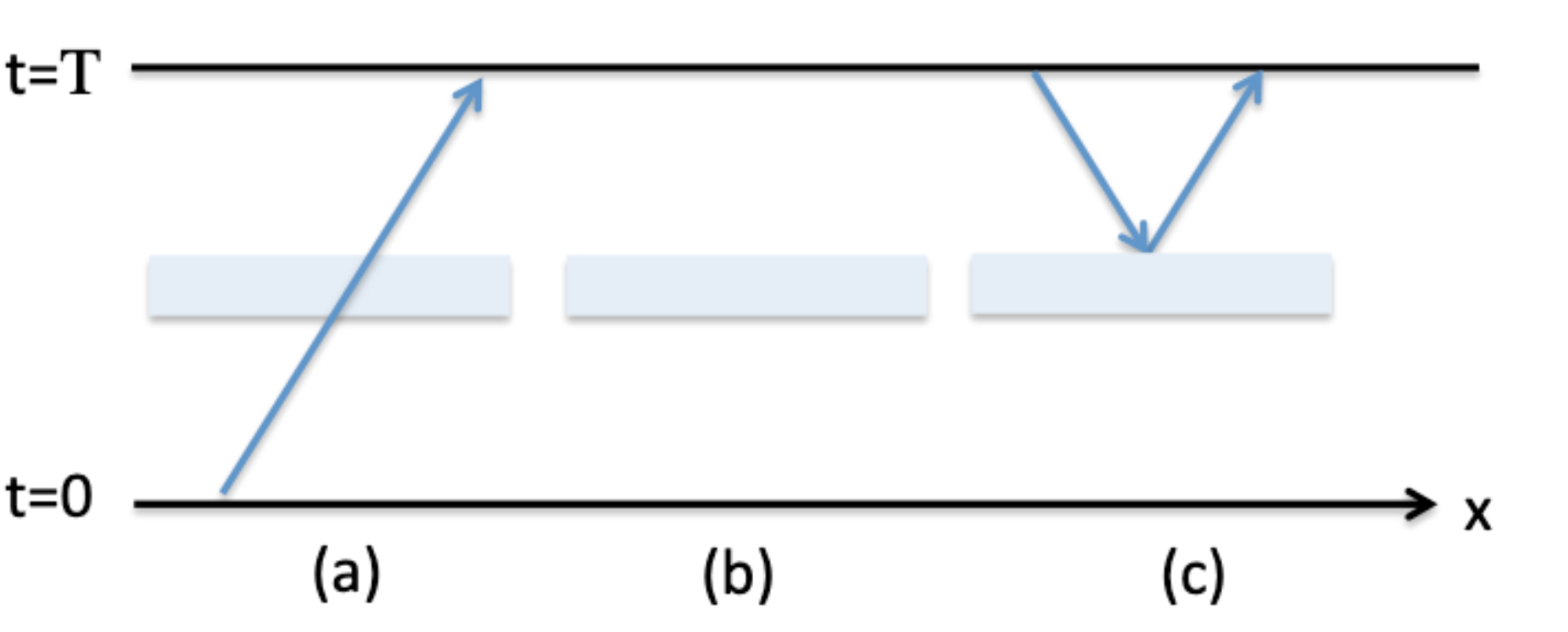}
    \caption{ The different scattering Feynman graphs involved in the interaction of a single mode $\mathbf{k}$ with the temporal Fabry Perot cavity. }
    \label{fig:7}
\end{figure}
\indent We start with a  simple example: Imagine we have initially in vacuum (at time $t=0$) a single electron bispinor $\psi_0(t=0,\mathbf{x}):=f(\mathbf{x})$ associated with $E>0$. This is a associated with a wave packet and the quantum state in the electron-hole theory is given by 
\begin{eqnarray}
|In \rangle=\hat{F}^\dagger|\varnothing\rangle
\end{eqnarray}   with $\hat{F}^\dagger=\int d^3\mathbf{x}\hat{\Psi}^\dagger(\mathbf{x})f(\mathbf{x})$ and $\hat{\Psi}^\dagger(\mathbf{x})=\sum_n \hat{a}^\dagger_nw_n(\mathbf{x})$, the sum is over positive and negative energy modes and we have $[\hat{a}_n,\hat{a}_m]_+=0$, $[\hat{a}_n,\hat{a}^\dagger_m]_+=\delta_{n,m}$. The probability amplitude to find the electron in the mode $g(\mathbf{x})$ at time $T$ is given by
\begin{eqnarray}
a=\langle\varnothing|\hat{G}\hat{U}(T,0)\hat{F}^\dagger|\varnothing\rangle
\end{eqnarray} with $\hat{G}=\int d^3\mathbf{x}g^\dagger(\mathbf{x})\hat{\Psi}(\mathbf{x})$. As shown by Feynman~\cite{Feynman1949} at time $t=T$ the wave $f(\mathbf{x})$ has evolved into $\psi_0(t=T,\mathbf{x}):=f'(\mathbf{x})$  and using conservation of the norm, the definition $\hat{\Psi}(\mathbf{x},T)=\hat{U}^{-1}(T,0)\hat{\Psi}(\mathbf{x})\hat{U}(T,0)$ (Heisenberg representation), and the new operator $\hat{F'}^\dagger=\int d^3\mathbf{x}\hat{\Psi}^\dagger(\mathbf{x})f'(\mathbf{x})$ we obtain 
\begin{eqnarray}
\hat{U}(T,0)\hat{F}^\dagger=\hat{F'}^\dagger\hat{U}(T,0).
\end{eqnarray} 
Moreover, in QED  we must modify  this expression~\cite{Feynman1949} since  a pure positive energy wave $f(\mathbf{x}):=f_{pos.}(\mathbf{x})$ should evolves as a sum containing both positive and negative  energy waves and these modes must be avoided. We solve this issue by writting the operator as 
\begin{eqnarray}
\hat{U}(T,0)(\hat{F}_{pos.}^\dagger+\hat{F}_{neg.}^\dagger)=\hat{F'}_{pos.}^\dagger\hat{U}(T,0).
\end{eqnarray}  where the operator $\hat{F}_{neg.}^\dagger)$ create a negative energy wave $f_{neg.}(\mathbf{x})$ in such a way that the sum $f_{pos.}(\mathbf{x})+f_{neg.}(\mathbf{x})$ evolves as a wave $f'_{pos.}(\mathbf{x})$ containing only positive energy at time $T$.  Therefore we obtain: 
\begin{eqnarray}
a=\langle\varnothing|\hat{G}(-\hat{U}(T,0)\hat{F}_{neg.}^\dagger+\hat{F'}_{pos.}^\dagger\hat{U}(T,0))|\varnothing\rangle.
\end{eqnarray} Moreover from the anticommutation rule we have also $\hat{G}\hat{F'}_{pos.}^\dagger=-\hat{F'}_{pos.}^\dagger\hat{G}+\int d^3\mathbf{x}g^\dagger(\mathbf{x})f'_{pos.}(\mathbf{x})$. Therefore, using the fact that the Dirac sea is full (all negative energy levels are occupied) we get
\begin{eqnarray}
a=\langle\varnothing|\hat{U}(T,0)|\varnothing\rangle \int d^3\mathbf{x}g^\dagger(\mathbf{x})f'_{pos.}(\mathbf{x})=C_v\int d^3\mathbf{x}g^\dagger(\mathbf{x})f'_{pos.}(\mathbf{x})
\end{eqnarray} where we introduced the vacuum to vacuum amplitude $C_v=\langle\varnothing|\hat{U}(T,0)|\varnothing\rangle $.    
In the case where the wave $g(\mathbf{x})$ is precisely one of the basis mode  $w_n(\mathbf{x})$ the amplitude $\int d^3\mathbf{x}w_n^\dagger(\mathbf{x})f'_{pos.}(\mathbf{x})$ is the expansion coefficient $C_n$ in the sum $\psi(t=T,\mathbf{x})=\sum_n C_n w_n(\mathbf{x})=\sum_nw_n(\mathbf{x})\int d^3\mathbf{x'}w_n^\dagger(\mathbf{x'})f'_{pos.}(\mathbf{x'})$ where $\psi(t=T,\mathbf{x})=\psi_0(t=T,\mathbf{x})+\int d^4x' S_F(x-x')e\gamma A(x')\psi(x')$ is the solution of Dirac's equation for the scattering process in the potential $A(x)$ using the Feynman propagator $S_F(x-x')$~\cite{Feynman1949}.\\
\indent As an application we consider the case of Fig.~\ref{fig:7}(a) where a positive energy plane wave is transmitted through the Fabry-Perot cavity of Appendix~\ref{Sec3}. In agreement with Eqs.~\ref{eq1tot} and \ref{eq2tot} an initial wave
\begin{eqnarray}
\psi_0(\mathbf{x},t=0):=f_{pos.}(\mathbf{x})=\sqrt{\frac{E+m}{2EV}}\left(\begin{array}{c}
  \chi\\ \frac{\boldsymbol{\sigma}\cdot\mathbf{p}}{E+m}\chi
  \end{array}\right)e^{i\mathbf{p}\cdot\mathbf{x}}
\end{eqnarray} is transmitted  and leads at time $t=T$: 
\begin{eqnarray}
\psi(\mathbf{x},t=T)=t_{tot.}(E,E')\sqrt{\frac{E+m}{2EV}}\left(\begin{array}{c}
  \chi\\ \frac{\boldsymbol{\sigma}\cdot\mathbf{p}}{E+m}\chi
  \end{array}\right)e^{i\mathbf{p}\cdot\mathbf{x}}e^{-iET}.
\end{eqnarray}
The probability amplitude  for the transmission  channel reads therefore  
\begin{eqnarray}
a=C_vt_{tot.}(E,E')e^{-iET}.
\end{eqnarray} and $C_v$ correponds to Fig.~\ref{fig:7}(b). The transition probability is thus given by   
\begin{eqnarray}
P_a=|a|^2=|C_v|^2|t_{tot.}(E,E')|^2=P_vT_{tot.}(E,E').
\end{eqnarray}
\indent A different  important  process is pair creation related to the transition amplitude\footnote{Note that the order of operators is arbitrary we could use instead $c'=\langle\varnothing|\hat{Q}^\dagger\hat{G}\hat{U}(T,0)|\varnothing\rangle$ but this will introduce a minus sign ($c'=-c$) since $\hat{Q}^\dagger\hat{G}=-\hat{G}\hat{Q}^\dagger+\int d^3\mathbf{x}g^\dagger(\mathbf{x})q(\mathbf{x})=-\hat{G}\hat{Q}^\dagger$ (we have $\int d^3\mathbf{x}g^\dagger(\mathbf{x})q(\mathbf{x}):=\int d^3\mathbf{x}g_{pos.}^\dagger(\mathbf{x})q'_{neg.}(\mathbf{x})=0$).} 
\begin{eqnarray}
c=\langle\varnothing|\hat{G}\hat{Q}^\dagger\hat{U}(T,0)|\varnothing\rangle
\end{eqnarray}
where $\hat{Q}^\dagger=\int d^3\mathbf{x}\hat{\Psi}^\dagger(\mathbf{x})q(\mathbf{x})$ is a creation operator for an electron with negative energy.  The  final state  $\hat{Q}\hat{G}^\dagger|\varnothing\rangle$ is thus associated with  the creation of a hole in the Dirac sea with $Q$ (i.e., the  annhilation of an electron with negative energy)  and the creation of an electron with positive energy with $\hat{G}^\dagger$. We calculate the amplitude $c$ with the same method as previously~\cite{Feynman1949}. We write $\hat{Q}^\dagger:=\hat{Q'}_{neg.}^\dagger=\int d^3\mathbf{x}\hat{\Psi}^\dagger(\mathbf{x})q'_{neg.}(\mathbf{x})$ and use the operator rule $\hat{U}(T,0)\hat{Q}_{neg.}^\dagger=(\hat{Q'}_{pos.}^\dagger+\hat{Q'}_{neg.}^\dagger)\hat{U}(T,0)$ that again corresponds to the Feynman Green function choice $S_F(x-x')$. After calculation we deduce:
\begin{eqnarray}
c=-C_v \int d^3\mathbf{x}g^\dagger(\mathbf{x})q'_{pos.}(\mathbf{x}),
\end{eqnarray} where this time $\int d^3\mathbf{x}g^\dagger(\mathbf{x})q'_{pos.}(\mathbf{x})$ is the coefficient amplitude for scattering of a negative energy electron $q'_{neg.}(\mathbf{x}):=\psi_0(T,\mathbf{x})$ at time $t=T$ into a positive energy particle at time $t=T$.  In the example of the temporal Fabry-Perot cavity  located at time $t_\mathcal{C}$ this leads to the Feynmlan graph of Fig.~\ref{fig:7}(c) and we obtain:
 \begin{eqnarray}
c=-C_vr'_{tot.}(E,E')e^{iET}e^{-iE(T-t_\mathcal{C})}e^{-iE(T-t_\mathcal{C})}=-C_vr'_{tot.}(E,E')e^{-iET}e^{2iEt_\mathcal{C}},
\end{eqnarray} 
and therefore the probability:
\begin{eqnarray}
P_c=|c|^2=|C_v|^2|r'_{tot.}(E,E')|^2=P_vR_{tot.}(E,E').
\end{eqnarray}
 Now since we have $P_v+P_c=1$, i.e., $P_v=1/(1+R_{tot.})$ this agrees with the result for \begin{eqnarray}
 P_a=P_vT_{tot.}=\frac{T_{tot.}}{1+R_{tot.}}=1.
\end{eqnarray}   Therefore, the Feynman procedure preserves probability conservation.\\
\indent An important property of this analysis (that is very general) concerns the number of modes used in the scattering process. Indeed,   we here considered  onle one fermion mode coresponding to a single wave-vector $\mathbf{k}$ associated with the incident electron. However, electron/positron pairs can be  extracted from the vacuum with any wavevector $\mathbf{k}$.  It is easy to see that the full probability for pair productions must obey
\begin{eqnarray}
1=P_v[1+R_{k_1}+R_{k_2}+R_{k_1}R_{k_2}+...]=\lim_{N\rightarrow+\infty}P_v\prod_{i=1}^{i=N}(1+R_{k_i}) \label{renor}
\end{eqnarray} where $R_{k_i}:=R_{tot.}(E_i,E_i')$ is the reflectivity associated with the mode  $\mathbf{k}_i$ (supposing a discrete basis).  Moreover, in our study  corresponding to Fig.~\ref{fig:7} the diagram (a) must be replaced by a diagram where pairs can be generated in all modes excluding the one   (i.e., $\mathbf{k}_1$) in which the electron is incident (i.e., to satisfy Pauli principle).   We are thus actually measuring the events with probability
\begin{eqnarray}
P_{a'}=\lim_{N\rightarrow+\infty}P_vT_{tot.}(E_1,E_1')\prod_{i=2}^{i=N}(1+R_{k_i})=\lim_{N\rightarrow+\infty}\frac{T_{tot.}(E_1,E_1')\prod_{i=2}^{i=N}(1+R_{k_i})}{\prod_{i=1}^{i=N}(1+R_{k_i})}=\frac{T_{tot.}(E_1,E_1')}{1+R_{tot.}(E_1,E_1')}=P_a.
\end{eqnarray} This is exactly the same result  as previously in the single mode calculation.
Of course to make mathematically sense of this calculation one must introduce a cut-off on the number of modes and in the end go to the limit $N\rightarrow + \infty$.    This method will be implicitely supposed in the following so that the not measured modes associated with unregistered pairs will be avoided from the discussion       
\subsection{First application}
\indent We now apply the previous method to the processes of Fig.~2 of the main paper and shown below in Fig.~\ref{fig:8}.
\begin{figure}[h]
    \centering
    \includegraphics[width=8cm]{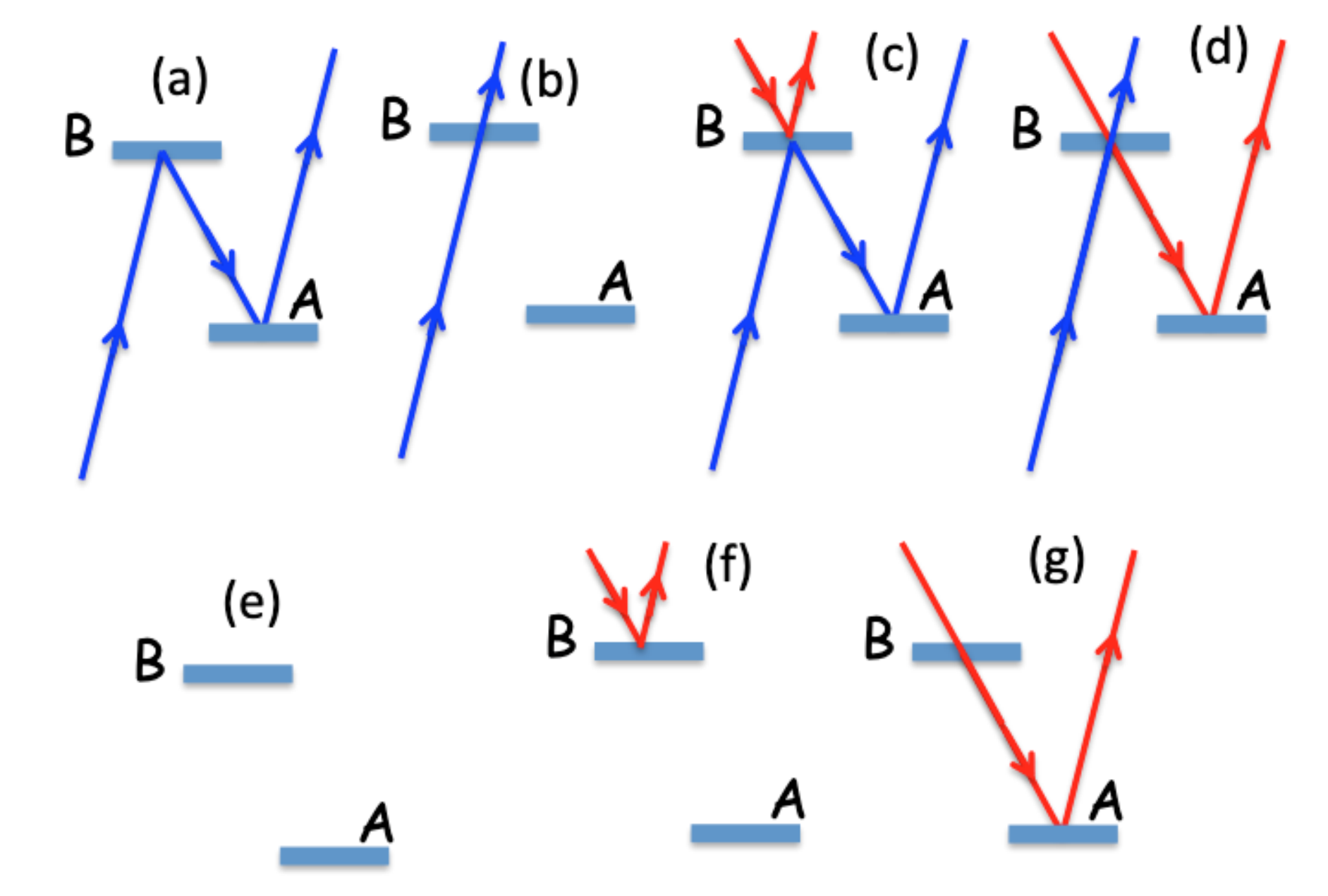}
    \caption{ (a-d) The different Feynman graphs involved in the scattering of a electron by a pair of temporal Fabry-Perot cavity (see text). (e-g) The different Feynman graphs involved if the initial state is the vacuum (i.e., Dirac sea).  }
    \label{fig:8}
\end{figure}
The Feynman `zigzag' diagram of Fig.~\ref{fig:8}(a) corresponds to a scattering amplitude
\begin{eqnarray}
a=C_vr'_{tot.}(E,E')r_{tot.}(E,E')e^{-iEt_\mathcal{B}}e^{-iE(t_\mathcal{B}-t_\mathcal{A})}e^{-iE(T-t_\mathcal{A})}\nonumber\\
=C_vr'_{tot.}(E,E')r_{tot.}(E,E')e^{-iET}e^{-i2E(t_\mathcal{B}-t_\mathcal{A})}
\end{eqnarray} and leads to the transition probability
  \begin{eqnarray}
P_a=|a|^2=P_vR'_{tot.}(E,E')R_{tot.}(E,E')=P_vR^2_{tot.}(E,E').
\end{eqnarray} In the same way Feynman's diagram of Fig.~\ref{fig:8}(b) leads to the probability:
\begin{eqnarray}
P_b=|b|^2=P_v|t_{tot.}(E,E')e^{-iET}|^2=P_vT_{tot.}(E,E').
\end{eqnarray} 
The two Feynman's diagrams of Figs.~\ref{fig:8}(c) and (d) must be considered together. Suppose that we have a scattering amplitude
\begin{eqnarray}
r=\langle\varnothing|\hat{G}_\alpha\hat{G}_\beta\hat{Q'}_{pos.}^\dagger\hat{U}(T,0)\hat{F}_{pos.}^\dagger|\varnothing\rangle
\end{eqnarray}  where a positive energy mode $f_{pos.}(\mathbf{x})$ at time $t=0$  evolves

\footnote{We stress that the typical mode  \begin{eqnarray}
\psi(\mathbf{x},t)=\sqrt{\frac{E+m}{2E}}\left(\begin{array}{c}
  \chi\\ \frac{\boldsymbol{\sigma}\cdot\mathbf{p}}{E+m}\chi
  \end{array}\right)\frac{e^{i\mathbf{p}\cdot\mathbf{x}}e^{-iEt}}{\sqrt{V}} \nonumber
\end{eqnarray} is now defined in a volume $V$ which is limited to the size of a beam ($V$ is not associated with an infinite box). The mode volume $V$ is interpreted as the spatial extension of a large quasi-monochromatic wave packet.} at time $t=T$ as two electrons with positive energies and modes $g_\alpha(\mathbf{x}),g_\beta(\mathbf{x})$ and one hole  associated with the destruction of the negative energy mode $q'_{neg.}(\mathbf{x})$. We obtain after applying the Feynman rules:
\begin{eqnarray}
r=C_v[\int d^3\mathbf{x}g_\alpha^\dagger(\mathbf{x})q'_{pos.}(\mathbf{x})\int d^3\mathbf{x}g_\beta^\dagger(\mathbf{x})f'_{pos.}(\mathbf{x})-\int d^3\mathbf{x}g_\beta^\dagger(\mathbf{x})q'_{pos.}(\mathbf{x})\int d^3\mathbf{x}g_\alpha^\dagger(\mathbf{x})f'_{pos.}(\mathbf{x})]
\end{eqnarray} where the minus sign is a consequence of the exclusion principle and is associated with a permutation in the final modes $g_\alpha\leftrightarrow g_\beta$.\\   
\indent Therefore, only the difference of the two  Feynman's diagrams of Figs.~\ref{fig:8}(c) and (d) has a physical meaning. 
Up to an arbitrary overall minus sign we get:
 \begin{eqnarray}
c-d=C_v[-r'_{tot.}(E,E')r_{tot.}(E,E')r'_{tot.}(E,E')e^{-iET}e^{-i2E(t_\mathcal{B}-t_\mathcal{A})}e^{-iET}e^{2iEt_\mathcal{B}}\nonumber\\
+t_{tot.}(E,E')t'_{tot.}(E,E')r'_{tot.}(E,E')e^{-iET}e^{-iET}e^{2iEt_\mathcal{A}}]\nonumber\\
=-C_v[r'_{tot.}(E,E')r_{tot.}(E,E')r'_{tot.}(E,E')-t_{tot.}(E,E')t'_{tot.}(E,E')r'_{tot.}(E,E')]e^{-i2E(T-t_\mathcal{A})}.
\end{eqnarray} This leads to the probability
\begin{eqnarray}
P_{c,d}=|c-d|^2=P_v|r'_{tot.}(E,E')r_{tot.}(E,E')r'_{tot.}(E,E')-t_{tot.}(E,E')t'_{tot.}(E,E')r'_{tot.}(E,E')|^2\nonumber\\
=P_vR_{tot.}(E,E')|r_{tot.}(E,E')r'_{tot.}(E,E')-t_{tot.}(E,E')t'_{tot.}(E,E')|^2=P_vR_{tot.}(E,E'),
\end{eqnarray} where we used Eq.~\ref{truc1} to obtain $|r_{tot.}(E,E')r'_{tot.}(E,E')-t_{tot.}(E,E')t'_{tot.}(E,E')|=1$.
 To complete the analysis of the system sketched in Fig.~\ref{fig:8} we must calculate the vacuum to vacuum probability $P_v:=P_e$ (see Fig.~\ref{fig:8}(e)).  This is done by considering the two diagrams of Figs.~\ref{fig:8}(f,g) corresponding to the creation of a pair at $\mathcal{A}$ or $\mathcal{B}$. We obtain
 \begin{eqnarray}
P_{f}=|f|^2=P_v|r'_{tot.}(E,E')|^2=P_vR_{tot.}(E,E')\nonumber\\
P_{g}=|g|^2=P_v|t'_{tot.}(E,E')r'_{tot.}(E,E')|^2=P_vT_{tot.}(E,E')R_{tot.}(E,E').\label{pair}
\end{eqnarray}  
Moreover, since we must have $P_e+P_f+P_g=1$ we deduce
\begin{eqnarray}
P_{v}=\frac{1}{(1+R_{tot.}(E,E'))^2}.\label{vide}
\end{eqnarray}  This shows that probability is also conserved  in channels associated with  Figs.~\ref{fig:8}(a-d)  since we have 
\begin{eqnarray}
P_a+P_b+P_{c,d}=P_v(1+R_{tot.}(E,E'))^2=1.
\end{eqnarray}
We mention that in the present examples  Eq.~\ref{renor} is replaced by \begin{eqnarray}
1=\lim_{N\rightarrow+\infty}P_v\prod_{i=1}^{i=N}(1+R_{k_i})^2 \label{renorb}
\end{eqnarray} and once again all the irrelevant  (unmeasured) modes are dropped out from the discussion.

\subsection{Second application}
\indent We now apply the previous method to the processes of Fig.~3(a) of the main paper and shown below in Fig.~\ref{fig:9}.
\begin{figure}[h]
    \centering
    \includegraphics[width=10cm]{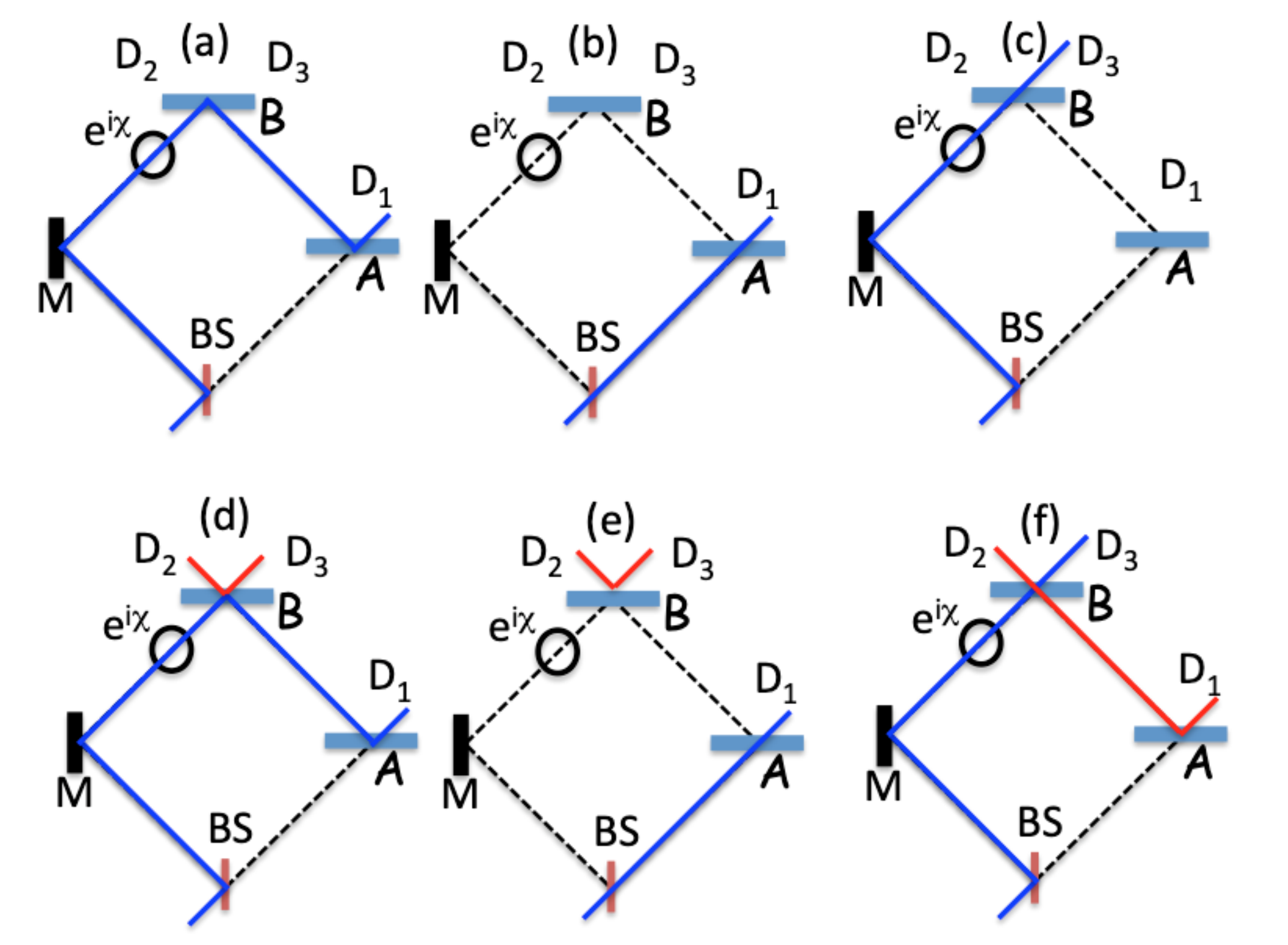}
    \caption{ Feynman graphs  associated with the interaction of a single electron with an interferometer involving temporal Fabry-Perot cavity. Interference with terms including pair creations are taken into account.   }
    \label{fig:9}
\end{figure}
In this interferometer we use a balanced beam splitter $BS$ with reflection amplitude $i/\sqrt{2}$ and transmission amplitude $1/\sqrt{2}$. The `zigzag' Feynman graph corresponding to Fig.~\ref{fig:9}(a) has a scattering amplitude
\begin{eqnarray}
a=\frac{i}{\sqrt{2}}C_vr'_{tot.}(E,E')r_{tot.}(E,E')e^{i\chi}e^{-iET}e^{-i2E(t_\mathcal{B}-t_\mathcal{A})}.
\end{eqnarray}This term interfere with the Feynman diagram of Fig.~\ref{fig:9}(b) with the scattering amplitude
\begin{eqnarray}
b=\frac{1}{\sqrt{2}}C_vt_{tot.}(E,E')e^{-iET}.
\end{eqnarray} The sum of these two diagrams gives the probability 
\begin{eqnarray}
P_{a,b}=|a+b|^2=\frac{P_v}{2}|ir'_{tot.}(E,E')r_{tot.}(E,E')e^{i\chi}e^{-i2E(t_\mathcal{B}-t_\mathcal{A})}+t_{tot.}(E,E')|^2.
\end{eqnarray} 
The graph of Fig.~\ref{fig:9}(c) leads to the probability
\begin{eqnarray}
P_{c}=|c|^2=\frac{P_v}{2}|it_{tot.}(E,E')e^{i\chi}e^{-iET}|^2=\frac{P_v}{2}T_{tot.}(E,E').
\end{eqnarray} 
We can calculate the probability to have a particle at gate $D_1$ and $D_3$ and an antiparticle at gate $D_2$ by adding the three Feynman's graphs of Figs.~\ref{fig:9}(d-f). More precisely we have the probability
 \begin{eqnarray}
P_{d,e,f}=|d+e-f|^2
=\frac{P_v}{2}|i(r'_{tot.}(E,E')r_{tot.}(E,E')r'_{tot.}(E,E')-t_{tot.}(E,E')t'_{tot.}(E,E')r'_{tot.}(E,E'))e^{i\chi}e^{-i2E(t_\mathcal{B}-t_\mathcal{A})}\nonumber\\+t_{tot.}(E,E')r'_{tot.}(E,E')|^2,\nonumber\\
\end{eqnarray}  where the minus sign takes into account Pauli's exclusion principle.\\
\indent Developping $P_{a,b}$ and $P_{d,e,f}$ we obtain
\begin{eqnarray}
P_{a,b}=\frac{P_v}{2}[1+R_{tot.}(E,E')+R^2_{tot.}(E,E')-it^\ast_{tot.}(E,E')r^2_{tot.}(E,E')e^{i\chi}e^{-i2E(t_\mathcal{B}-t_\mathcal{A})}+cc.],\nonumber\\
P_{d,e,f}=\frac{P_vR_{tot.}(E,E')}{2}[2+R_{tot.}(E,E')-it'_{tot.}(E,E')e^{i\chi}e^{-i2E(t_\mathcal{B}-t_\mathcal{A})}+cc.]
\end{eqnarray} 
Moreover, from Eq.~\ref{truc1} we can easily deduce $R_{tot.}(E,E')t'_{tot.}(E,E')=-t^\ast_{tot.}(E,E')r^2_{tot.}(E,E')$ and consequently the oscillating terms in $P_{a,b}$ and $P_{d,e,f}$ cancel each other. After some manipulations we can directly obtain Eqs.~8, 9 of the main article. From this we obtain the probability conservation
\begin{eqnarray}
P_{a,b}+P_{c}+P_{d,e,f}=P_v(1+R_{tot.}(E,E'))^2.
\end{eqnarray}   
Importantly, as in the previous subsection we can evaluate $P_v$ by considering a pair creation at  $\mathcal{A}$ or $\mathcal{B}$. The result is the same as in Eq.~\ref{pair} and therefore we recover Eq.~\ref{vide}, i.e., $P_{v}=\frac{1}{(1+R_{tot.}(E,E'))^2}$. This in turn implies  
\begin{eqnarray}
P_{a,b}+P_{c}+P_{d,e,f}=1.
\end{eqnarray} The rule is in fact general $P_v$ can not be affected by the presence of devices like $BS$  or the phase shifter $e^{i\chi}$ since these are located before the pair creation at $\mathcal{A}$ or $\mathcal{B}$ as it is seen from the Feynman diagrams.\\
\indent From the previous results we have
\begin{eqnarray}
P(D_1,\varnothing D_{2,3})=\frac{P_v}{2}(T_{tot.}(E,E')+R_{tot.}(E,E')^2)(1+\mathcal{V}\sin{\theta});\nonumber\\
P(D_1,D_{2},D_3)=\frac{P_v}{2}(2R_{tot.}(E,E')+R_{tot.}(E,E')^2)-\frac{P_v}{2}(T_{tot.}(E,E')+R_{tot.}(E,E')^2)\mathcal{V}\sin{\theta},\nonumber\\
P(D_3,\varnothing D_{1,2})=P_v(1+R_{tot.}(E,E'))/2,\label{game}
\end{eqnarray}  with 
\begin{eqnarray}
\mathcal{V}=\frac{2R_{tot.}(E,E')\sqrt{T_{tot.}(E,E')}}{T_{tot.}(E,E')+R_{tot.}(E,E')^2}\leq 1,\nonumber\\
\theta=\chi-2E(t_\mathcal{B}-t_\mathcal{A})+2\arg{[r_{tot.}(E,E')]}-\arg{[t_{tot.}(E,E')]}.
\end{eqnarray} The constraint $\mathcal{V}\leq 1$ is directly obtained from the property $1+R_{tot.}(E,E')=T_{tot.}(E,E')$. The limit $\mathcal{V}=1$ is obtained if $0=1+R_{tot.}(E,E')-R^2_{tot.}(E,E')$ that admits the solution 
\begin{eqnarray}
R_{tot.}(E,E'):=R_0=\frac{1+\sqrt{5}}{2}\simeq 1.62.
\end{eqnarray} 
\section{Appendix 5: A simple retrocausal game (see Fig. 3(a) of the manuscript)}\label{Sec5}
\indent From Eq.~\ref{game} we consider the two cases $\theta_\pm=\pm\frac{\pi}{2}$, $\mathcal{V}=1$ (i.e., $R_{tot.}(E,E'):=R_0=\frac{1+\sqrt{5}}{2}$). This implies:
 \begin{eqnarray}
P(D_1,\varnothing D_{2,3}|\theta_-)=0;\nonumber\\
P(D_1,D_{2},D_3|\theta_-)=\frac{P_v}{2}(1+3R_{tot.}(E,E')+2R_{tot.}(E,E')^2),\nonumber\\
P(D_3,\varnothing D_{1,2}|\theta_-)=\frac{P_v}{2}(1+R_{tot.}(E,E')),\label{game1}
\end{eqnarray}
 and\begin{eqnarray}
P(D_1,\varnothing D_{2,3}|\theta_+)=P_v(1+R_{tot.}(E,E')+R_{tot.}(E,E')^2);\nonumber\\
P(D_1,D_{2},D_3|\theta_+)=\frac{P_v}{2}(R_{tot.}(E,E')-1),\nonumber\\
P(D_3,\varnothing D_{1,2}|\theta_+)=\frac{P_v}{2}(1+R_{tot.}(E,E')).\label{game2}
\end{eqnarray}
In our game we assume $P(\theta_\pm)=\frac{1}{2}$, i.e., that $\theta_\pm$ is selected by a fair quantum coin tossing. The guess $a$ of the value $\theta$ made by the agent is defined by the following rules:\\
\indent 1) If a single electron  is detected  at $D_1$ and none at $D_{2,3}$ the agent knows with certainty that $\theta_+$ has been selected. This allows us to define  the probability for guessing $a$: 
\begin{eqnarray}
P(a,D_1,\varnothing D_{2,3};\theta_+)=\delta_{a,\theta_+}P(D_1,\varnothing D_{2,3}|\theta_+)P(\theta_+)=\delta_{a,\theta_+}\frac{P_v}{2}(1+R_0+R_0^2).\label{game3}
\end{eqnarray} Similarly we have 
\begin{eqnarray}
P(a,D_1,\varnothing D_{2,3};\theta_-)=0.\label{game3b}
\end{eqnarray} 
\\
\indent 2) If a single electron is detected at $D_3$ and none in $D_{1,2}$ or if all detectors $D_1,D_2$ and $D_3$ found a particle the agent has no way from the data to  unambiguously infer the value  $\theta$. The best she can do is to toss a fair quantum toss with probability $P=\frac{1}{2}$ for each values of $a$. This allows us to define the probabilities for guessing $a$:
  \begin{eqnarray}
P(a,D_3,\varnothing D_{1,2};\theta_\pm)=\frac{1}{2}P(D_3,\varnothing D_{1,2}|\theta_\pm)P(\theta_\pm)=\frac{P_v}{8}(1+R_0),\nonumber\\
P(a,D_1,D_{2},D_3;\theta_-)=\frac{1}{2}P(D_1,D_{2},D_3|\theta_-)P(\theta_-)=\frac{P_v}{8}(1+3R_0+2R_0^2),\nonumber\\
P(a,D_1,D_{2},D_3;\theta_+)=\frac{1}{2}P(D_1,D_{2},D_3|\theta_+)P(\theta_+)=\frac{P_v}{8}(R_0-1).
\label{game4}
\end{eqnarray} The sum of all probabilities in Eqs.~\ref{game3},\ref{game4} leads to 
\begin{eqnarray}
\sum_{a,\theta}P(a,D_1,\varnothing D_{2,3};\theta)+P(a,D_3,\varnothing D_{1,2};\theta)+P(a,D_1,D_{2},D_3;\theta)=1.
\end{eqnarray}
\indent In the game, inspired by  previous works \cite{Mafalda,Oreshkov2012,Branciard2016}, we define the mean gain of the agent by 
\begin{eqnarray}
\langle G\rangle=\sum_{a,\theta}\delta_{a,\theta}P(a,\theta)=\sum_{\theta}P(a=\theta,\theta)\nonumber\\
=\sum_{\theta}P(\theta,D_1,\varnothing D_{2,3};\theta)+P(\theta,D_3,\varnothing D_{1,2};\theta)+P(\theta,D_1,D_{2},D_3;\theta)
\end{eqnarray}
Moreover, by using the rules 1 and 2 and Eqs.~\ref{game3},\ref{game4} we deduce:
\begin{eqnarray}
\langle G\rangle_{quant.}=\frac{3}{4}-\frac{R_0}{4(1+R_0)^2}\simeq 0.69>1/2.
\end{eqnarray}
This value violates the classical bound obtained by an agent who not having access to a retrocausal channel cannot infer the probabilities  $P(a,\theta)$ from the data at detectors $D_1,D_2,D_3$.   Indeed in this classical regime changing $\chi$ and thus $\theta$ cannot influence the probabilities at $D_i$.  The only possibility for the agent is thus to use a fair coin (or some other means)  to guess the value $a$. Writing $P_{class.}(a,\theta)=P_{class.}(a|\theta)P_{class.}(\theta)$ with $P_{class.}(a|\theta)\leq \frac{1}{2}$ and with $P_{class.}(\theta)=\frac{1}{2}$ we obtain the bound: 
\begin{eqnarray}
\langle G\rangle_{class.}=\sum_{a,\theta}\delta_{a,\theta}P_{class.}(a,\theta)\leq \frac{1}{2}.
\end{eqnarray}
To increase the gain in the quantum game we can postselect on those cases where single electron  is detected  at $D_1$ and none at $D_{2,3}$ this leads to 
\begin{eqnarray}
\langle G\rangle_{postselected}=\sum_{\theta}P(\theta,D_1,\varnothing D_{2,3};\theta)=P(D_1,\varnothing D_{2,3};\theta_+)=\frac{P_v}{2}(1+R_0+R_0^2).
\end{eqnarray}Moreover, from the definition we can also  introduce the conditional probabilities
\begin{eqnarray}
P(\theta_\pm|D_1,\varnothing D_{2,3})=\frac{P(D_1,\varnothing D_{2,3};\theta_\pm)}{ P(D_1,\varnothing D_{2,3})}=\frac{P(D_1,\varnothing D_{2,3};\theta_\pm)}{\sum_\theta P(D_1,\varnothing D_{2,3};\theta)}=\frac{P(D_1,\varnothing D_{2,3}|\theta_\pm)}{\sum_\theta P(D_1,\varnothing D_{2,3}|\theta)P(\theta)}
\end{eqnarray} and from Eqs.~\ref{game3},\ref{game3b} we have $P(\theta_+|D_1,\varnothing D_{2,3})=1$ and $P(\theta_-|D_1,\varnothing D_{2,3})=0$. Therefore we can define the conditional gain as  \begin{eqnarray}
\langle G\rangle_{|D_1,\varnothing D_{2,3}}=\frac{\langle G\rangle_{postselected}}{P(D_1,\varnothing D_{2,3})}=\sum_{\theta}P(\theta|D_1,\varnothing D_{2,3})=1
\end{eqnarray} that is again violating the (conditional) classical bound defined this time as:
\begin{eqnarray}
\langle G\rangle_{class.|D_1,\varnothing D_{2,3}}=\frac{\sum_{\theta}P_{class.}(\theta,D_1,\varnothing D_{2,3};\theta)}{P_{class.}(D_1,\varnothing D_{2,3})}\leq \frac{\sum_{\theta}\frac{1}{2}P_{class.}(D_1,\varnothing D_{2,3};\theta)}{P_{class.}(D_1,\varnothing D_{2,3})}=\frac{1}{2}.
\end{eqnarray}
\section{Appendix 6: A retrocausal version of the quantum switch (see Fig. 3(b) of the manuscript)}\label{Sec6}
\indent We need to evaluate the vacuum to vacuum transition probability $P_v=|C_v|^2$ in presence of the interferometer of Fig. 3(b) of the manuscript. As a very general rule observe that $C_v=\langle\varnothing|\hat{U}(T,0)|\varnothing\rangle$ cannot depend on what happens at times before   $t_\mathcal{A}$.   The reason is that no source for pair creation are present  before  $t_\mathcal{A}$ (i.e., $\hat{U}(t_\mathcal{A},0)|\varnothing\rangle=e^{i\phi}|\varnothing\rangle$, with $\phi$ a irrelevant phase).  We can thus write (up to a phase factor) $C_v=\langle\varnothing|\hat{U}(T,t_\mathcal{A})|\varnothing\rangle$.  Moreover after $t_\mathcal{B}$ there is no source for pair creation and we can write $\langle\varnothing|\hat{U}(T,t_\mathcal{B})=e^{i\phi'}\langle\varnothing|$.  We have thus  $P_v=|\langle\varnothing|\hat{U}(t_\mathcal{B},t_\mathcal{A})|\varnothing\rangle|^2$. 
\begin{figure}[h]
    \centering
    \includegraphics[width=9cm]{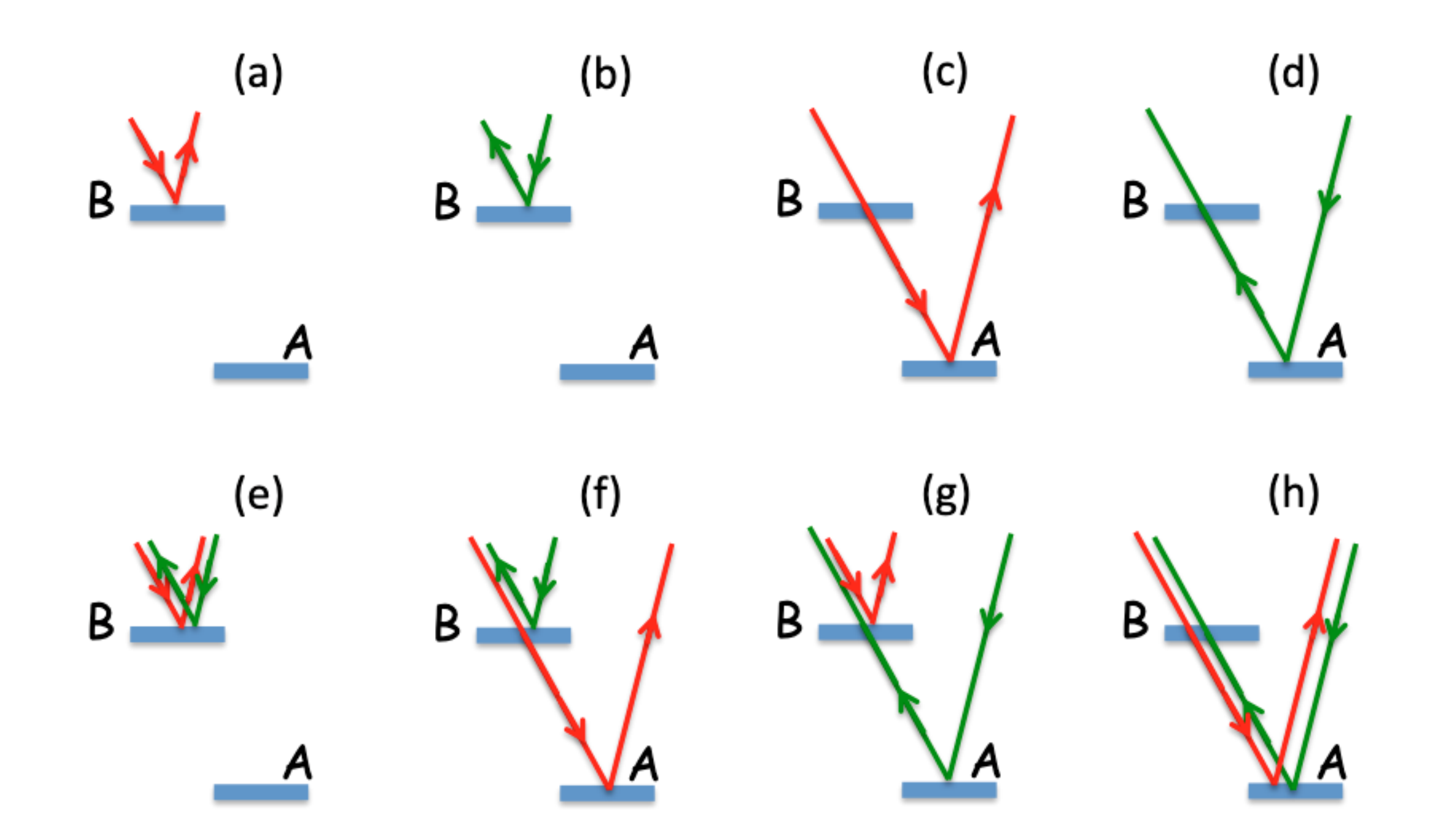}
    \caption{  Pair creation in two modes (red and green lines) corresponding to two wave vectors $\pm\mathbf{k}$ in presence of two temporal cavities.  }
    \label{fig:10}
\end{figure}
This corresponds to the physical situation depicted in Fig.~\ref{fig:10} where all the Feynman graphs involved in pair creations in the two cavities $\mathcal{A,B}$ are represented.  Comparing this situation to the case of Fig.~\ref{fig:8} we see that we now need to kinds of modes corresponding to wavevectors $\pm\mathbf{k}$. This is because in the interferometer of Fig. 3(b) of the manuscript we have a direct path from cavity $\mathcal{A}$ to cavity $\mathbf{B}$, and a zigzag path from cavity $\mathcal{B}$ to cavity $\mathbf{A}$.\\
\indent   The probabilities of the graphs $a-h$ are easily obtained:
\begin{eqnarray}
P_a=|a|^2=P_b=|b|^2=P_vR_0, &P_c=|c|^2=P_d=|d|^2=P_vR_0T_0, \nonumber\\
P_e=|e|^2=P_vR_0^2, &P_f=|f|^2=P_g=|g|^2=P_vR_0^2T_0,\nonumber\\
P_h=|h|^2=P_vR_0^2T_0^2,
\end{eqnarray}  where for simplicity we have written $R_{tot.}(E,E')=R_0$,  $T_{tot.}(E,E')=T_0=1+R_0$    and assumed that the two wavevectors $\pm\mathbf{k}$ have the same Fresnel coefficients due to symmetry. By adding all these probabilities to $P_v$ we have
\begin{eqnarray}
P_v+\sum_i P_i=P_v(1+R_0)^4=1,
\end{eqnarray} that defines $P_v$. Moreover, this time Eq.~\ref{renor} is replaced by \begin{eqnarray}
1=\lim_{N\rightarrow+\infty}P_v\prod_{i=1}^{i=N}(1+R_{k_i})^4. \label{renorc}
\end{eqnarray} \\
\indent For the present purpose we are mainly interested in calculating the amplitudes for a single electron to reach the detector $D_2$ or $D_3$.  The other terms are not relevant for the present study. For this purpose we start with a bispinor state before the beam splitter $BS_1$ of  Fig. 3(b) in the main article:  \begin{eqnarray}
\psi_{0}(\mathbf{p},E)\frac{e^{i\mathbf{p}\cdot\mathbf{x}}e^{-iEt}}{\sqrt{V}}=\sqrt{\frac{E+m}{2E}}\left(\begin{array}{c}
  \chi^{(0)}\\ \frac{\boldsymbol{\sigma}\cdot\mathbf{p}}{E+m}\chi^{(0)}
  \end{array}\right)\frac{e^{i\mathbf{p}\cdot\mathbf{x}}e^{-iEt}}{\sqrt{V}} 
\end{eqnarray} where $\chi^{(0)}$ is a spinor and $E$ is a positive energy.  In the path $BS_1\prec M_2\prec \mathcal{A}\prec \mathcal{B}\prec M_3\prec D_{2,3}$ the bispinor is first after the mirror  and the cavity $\mathcal{A}$ transformed into a $\psi_0(-\mathbf{p},E)\frac{e^{-i\mathbf{p}\cdot\mathbf{x}}e^{-iEt}}{\sqrt{V}}$ mode. The two unitaries $\hat{U}_A,\hat{U}_B$ acting on $\psi_0(-\mathbf{p},E)$ lead to 
$$\hat{U}_B\hat{U}_A\psi_0(-\mathbf{p},E)=\sum_s\psi_s(-\mathbf{p},E)C_s^{\mathcal{A}\prec\mathcal{B}}=\sum_s\psi_s(-\mathbf{p},E)\psi^\dagger_s(-\mathbf{p},E)\hat{U}_B\hat{U}_A\psi_0(-\mathbf{p},E)$$ where the basis states  $\psi_s$ ($s=\pm \frac{1}{2}$) are defined as $\psi_s(\mathbf{p},E)=\sqrt{\frac{E+m}{2E}}\left(\begin{array}{c}
  \chi_s\\ \frac{\boldsymbol{\sigma}\cdot\mathbf{p}}{E+m}\chi_s
  \end{array}\right)$  with $\chi_{+\frac{1}{2}}=\left(\begin{array}{c}
  1\\ 0
  \end{array}\right)$ and $\chi_{-\frac{1}{2}}=\left(\begin{array}{c}
  0\\ 1
  \end{array}\right)$. The coefficient $C_s^{\mathcal{A}\prec\mathcal{B}}=\psi^\dagger_s(-\mathbf{p},E)\hat{U}_B\hat{U}_A\psi^{(0)}_{-\mathbf{p},E}$ is a fundamental scattering matrix  for this  path $BS_1\prec M_2\prec \mathcal{A}\prec \mathcal{B}\prec M_3\prec D_{2,3}$ and this will preserved until the detectors $D_{2,3}$.   Same, in the the zigzag path $BS_1\prec M_1\prec \mathcal{B}\prec\mathcal{A}\prec M_4\prec D_{2,3}$ the initial state $\psi_{0}(\mathbf{p},E)$ is after the cavity $\mathcal{B}$ transformed into a negative energy state $\psi_{0}(\mathbf{p},-E)$ 
  and after the unitaries we obtain $$\hat{U}^{-1}_A\hat{U}^{-1}_B\psi_0(\mathbf{p},-E)=\sum_s\psi_s(\mathbf{p},-E)C_s^{\mathcal{B}\prec\mathcal{A}}=\sum_s\psi_s(\mathbf{p},-E)\psi^\dagger_s(\mathbf{p},-E)\hat{U}^{-1}_A\hat{U}^{-1}_B\psi_0(\mathbf{p},-E).$$ Importantly, after the cavity $\mathcal{A}$ the negative wave $\psi_s(\mathbf{p},-E)$ transforms into  $\psi_s(\mathbf{p},E)$ and the coeffcient $C_s^{\mathcal{B}\prec\mathcal{A}}$ is now associated with a positive energy state propagating forward in time. We stress that along the zigzag path the unitaries acting backward in time are  $\hat{U}^{-1}_{A,B}$ and not $\hat{U}_{A,B}$ (this is logical since an evolution operator 
$\hat{U}(\delta t)=e^{-iH\delta t}$, with $H$ an Hermitian operator,  is for a negative time delay $\delta t=-|\delta t|$ equivalent to $\hat{U}(-|\delta t])=\hat{U}^{-1}(+|\delta t])$).\\   
\indent It is straigthforward to obtain for the amplitudes $a_{D_2},a_{D_3}$:
\begin{eqnarray}
a_{D_2,s}=C_ve^{-iET}[\frac{1}{\sqrt{2}}t_{tot.}(E,E')t_{tot.}(E,E')\frac{i}{\sqrt{2}}C_s^{\mathcal{A}\prec\mathcal{B}}+e^{i\xi}\frac{i}{\sqrt{2}}r_{tot.}(E,E')r'_{tot.}(E,E')\frac{1}{\sqrt{2}}C_s^{\mathcal{B}\prec\mathcal{A}}]\nonumber\\
a_{D_3,s}=C_ve^{-iET}[\frac{1}{\sqrt{2}}t_{tot.}(E,E')t_{tot.}(E,E')\frac{1}{\sqrt{2}}C_s^{\mathcal{A}\prec\mathcal{B}}+e^{i\xi}\frac{i}{\sqrt{2}}r_{tot.}(E,E')r'_{tot.}(E,E')\frac{i}{\sqrt{2}}C_s^{\mathcal{A}\prec\mathcal{B}}]\label{proba}
\end{eqnarray} 
where $\xi=\chi-2E(t_\mathcal{B}-t_\mathcal{A})$. Eq.~\ref{proba} directly implies Eq.~13 of the main article~\footnote{The norm $\lVert \rVert^2$ } with $P(D_2,\varnothing D_{3,1})=\sum_s\lVert a_{D_2,s}\rVert^2$ and  $P(D_3,\varnothing D_{2,1})=\sum_s\lVert a_{D_3,s}\rVert ^2$.
We have thus: 
\begin{eqnarray}
P(D_2,\varnothing D_{3,1})=\frac{P'_v}{4}\sum_s\lVert t^2_{tot.}(E,E')C_s^{\mathcal{A}\prec\mathcal{B}}-r^2_{tot.}(E,E')e^{i\xi}C_s^{\mathcal{B}\prec\mathcal{A}}\rVert ^2,\nonumber\\
P(D_3,\varnothing D_{2,1})=\frac{P'_v}{4}\sum_s\lVert t^2_{tot.}(E,E')C_s^{\mathcal{A}\prec\mathcal{B}}+r^2_{tot.}(E,E')e^{i\xi}C_s^{\mathcal{B}\prec\mathcal{A}}\rVert ^2\label{ICO}
\end{eqnarray} 
\section{Appendix 7: Closed time like curves}\label{Sec7}
\indent Closed time like curves (CTCs) have been discussed in the context of general relativity and are subject to self-consistency paradoxes like the so called `grand father paradox' in which a time traveller changing the past prohibits her own existence  and  therefore leads to a contradiction.  
\begin{figure}[h]
    \centering
    \includegraphics[width=9cm]{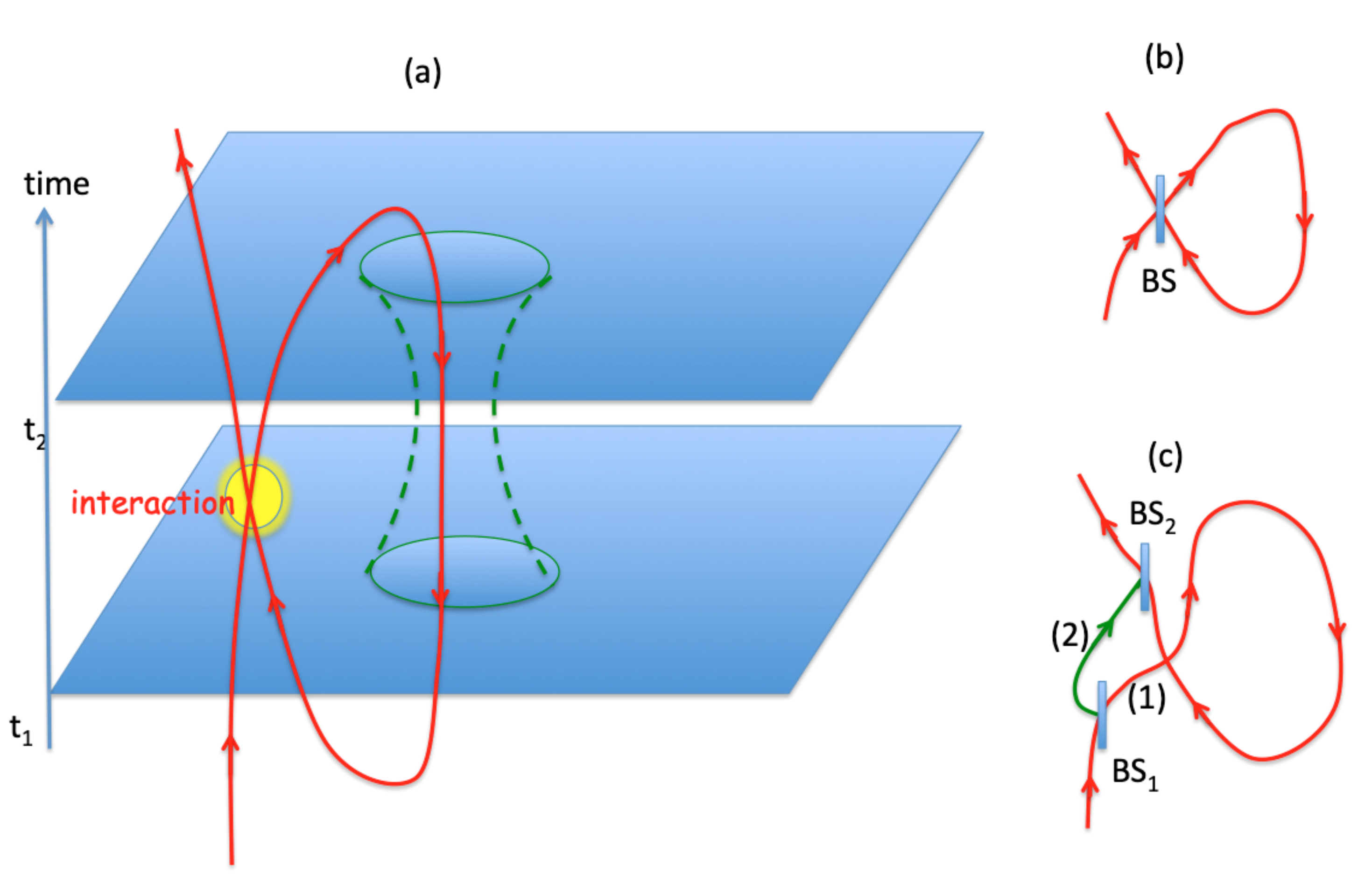}
    \caption{(a) Closed time-like curves  (CTCs) in presence of a gravitational wormhole acting as a time machine \cite{Thorne1}. (b-c) circuit analogies for two different CTCs solving the grand-father paradox. }
    \label{fig:11}
\end{figure}
In the Polchinski collision paradox~\cite{Echeverria} (see Fig.~\ref{fig:11}(a)) a particle entering into a gravitational wormhole acting as a time machine \cite{Thorne1,Thorne2} can arrive into the past and collides with itself prohibiting the particle to enter the CTC. Moreover, in quantum mechanics this paradox can be trated into two different ways.  In the first one based on a consistency condition we consider the time travel of a qunatum state into its ow past as a kind of temporal ring interferometer  or cavity. As sketched in  Fig.~\ref{fig:11}(b) a single particle wave function incident on a beam splitter can either be transmitted or reflected.  The transmitted wave is sent through the wormhole and after emerging into the past interferes with itself on the beam splitter (mimicking the Polchinski collision). The presence of a CTC doesn't however lead to any paradox \cite{Greenberger} and can be analyzed as any closed loop interferometer. In the second strategy due to D. Deutsch~\cite{Deutsch} we avoid the grandfather paradox by allowing the possibitly to interact with a parallel branch of the wave function.   A simple illustration is shown in Fig.~\ref{fig:11}(c): A particle wave function first interacts with a beam splitter dividing it into to branches 1 and 2. Wave 1 goes through the time machine  but when it is going back into the past it doesnt interact with itself  but instead with beam 2 at a second beam splitter.  Nothing prohibits us to tune the phase in order to have a complete constructive interference in one exit and destructive in the other one. Therefore, the wave packet could be deviated faraway from the time machine and will not enter into a CTC.  This again avoids any paradox.\\
\begin{figure}[h]
    \centering
    \includegraphics[width=9cm]{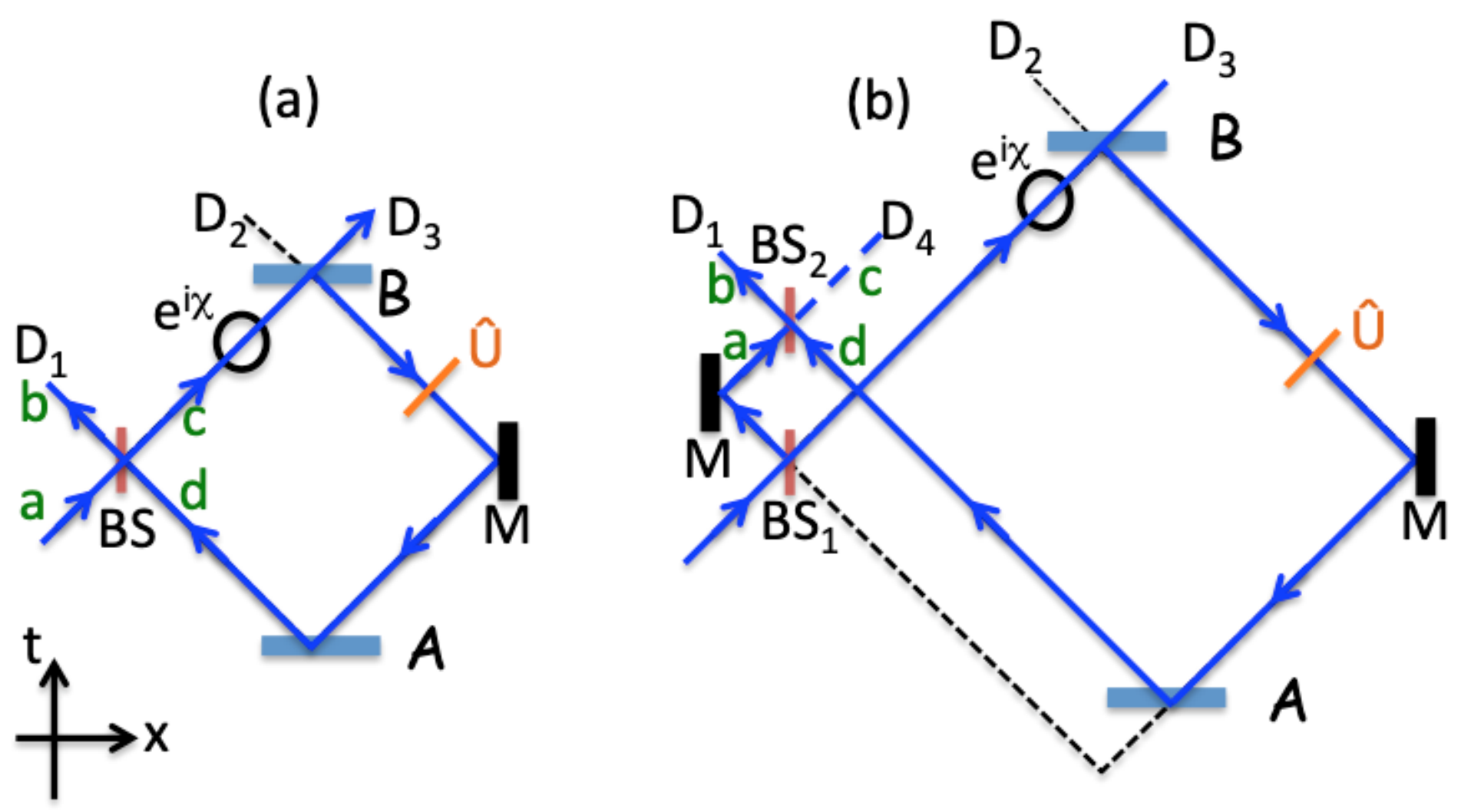}
    \caption{ Implementations of time machine and self-consistent CTCS using temporal Fabry-Perot cavities for single electron allowed to move backward in time.}
    \label{fig:12}
\end{figure}
\indent In the context of our work we can develop examples of CTCs free from contradiction using either the Fig.~\ref{fig:11}(b)  or (c) strategies.\\ 

\begin{figure}[h]
    \centering
    \includegraphics[width=9cm]{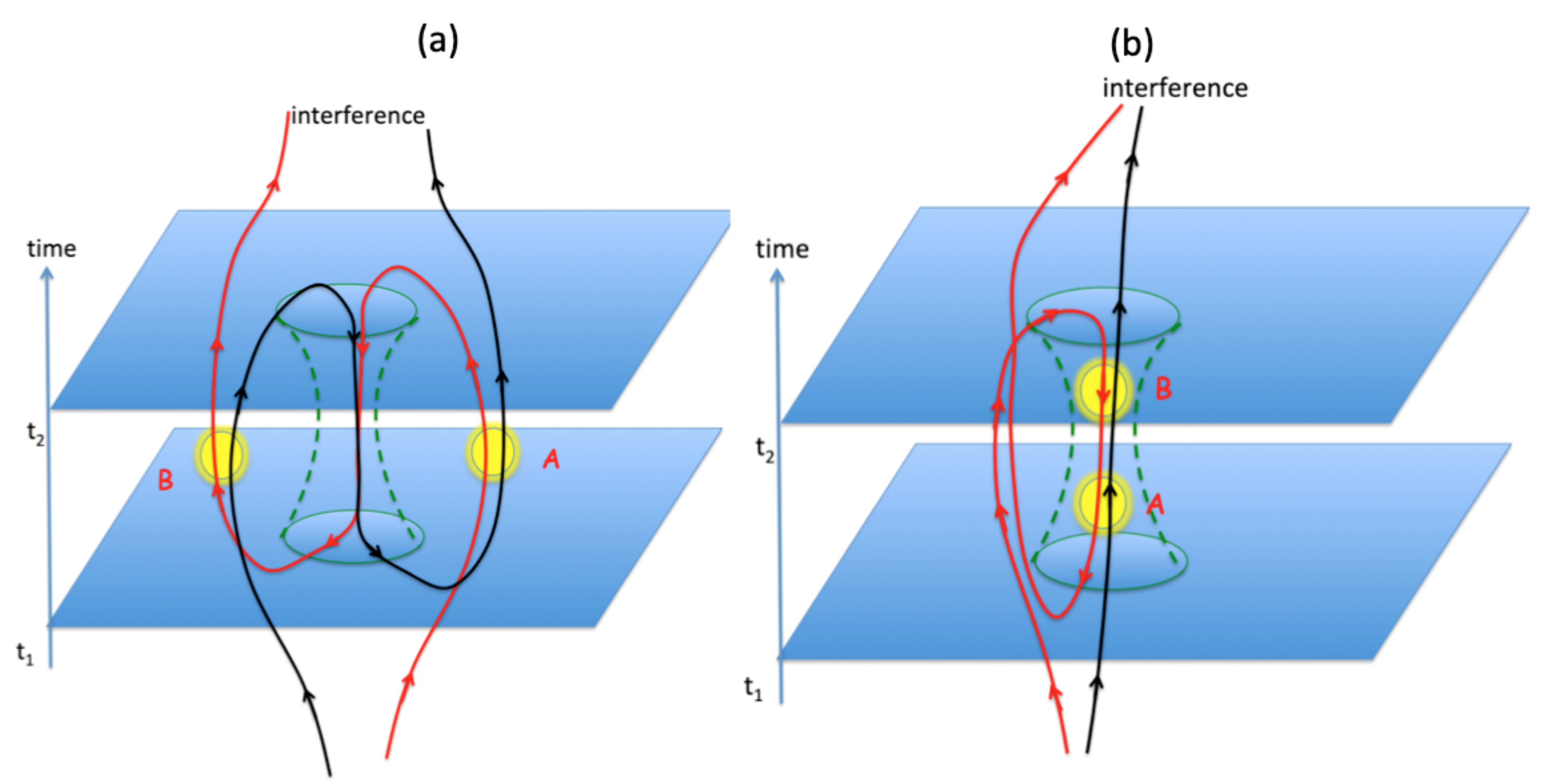}
    \caption{Two examples of gravitational time machines inspired of \cite{Thorne2} and used for implementing quantum switches 	and indefinite causal order.   }
    \label{fig:13}
\end{figure}

\begin{figure}[h]
    \centering
    \includegraphics[width=9cm]{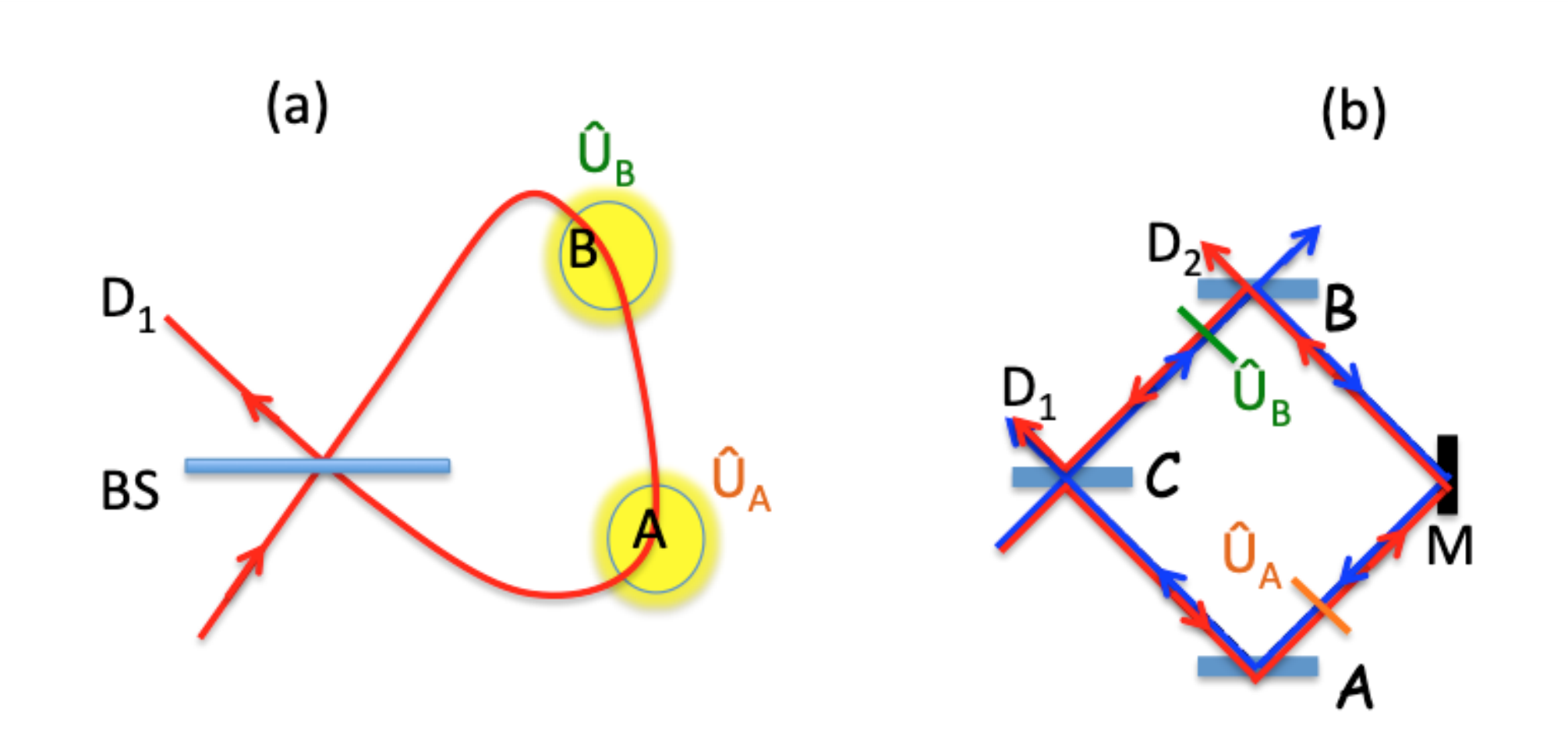}
    \caption{(a) the simplest optical QS implementation involving only one beam splitter and two unitaries acting on the polarization state of the photon (the photon is detected in gate $D_1$ and can move in the cavity using two possible counterpropagating paths (note that here part of the photon state also exits in the incident path).  (b) A relativistic version involving CTCs and three temporal Fabry-Perot cavities $\mathcal{A,B,C}$.  The unitaries   $\hat{U}_{A,B}$ act on the internal (polarization, spin) state of the particle. }
    \label{fig:14}
\end{figure}

\indent 1) In the first strategy, illustrated in Fig.~\ref{fig:12}(a), we use two temporal cavities $\mathcal{A,B}$, a mirror and a 50/50 beam splitter $BS$. The sytem acts as a ring resonator for a single electron making a CTC and we included a phase shift $e^{i\chi}$ and a unitary $\hat{U}$ acting on the bi-spinor state of the single electron. The entrance and exit gates of BS are written $a,d$ and $b,c$.    The input is in $a$ and we measure the electron in  beam $b$ where a detector $D_1$ is inserted. As usual other detectors $D_{2,3}$ are included in order exits since they play a role in the evaluation of $P_v$ and because all our analysis requires postselections based on correlations.\\
\indent From the properties of the 50/50 $BS$ we write for the four bispinors  $\psi_a,\psi_b,\psi_c,\psi_d$  in the channels $a,b,c,d$:
\begin{eqnarray}
\psi_a=\sum_s a_s\psi_s(\mathbf{p},E),&\psi_c=\sum_s c_s\psi_s(\mathbf{p},E)\nonumber\\
\psi_b=\sum_s b_s\psi_s(-\mathbf{p},E),&\psi_d=\sum_s c_s\psi_s(-\mathbf{p},E)
\end{eqnarray} with the unit spinor basis $\psi_s(\mathbf{p},E)=\sqrt{\frac{E+m}{2E}}\left(\begin{array}{c}
  \chi_s\\ \frac{\boldsymbol{\sigma}\cdot\mathbf{p}}{E+m}\chi_s
  \end{array}\right)$ with with $\chi_{+\frac{1}{2}}=\left(\begin{array}{c}
  1\\ 0
  \end{array}\right)$ and $\chi_{-\frac{1}{2}}=\left(\begin{array}{c}
  0\\ 1
  \end{array}\right)$. The amplitudes are obeying the unitary conditions 
\begin{eqnarray}
c_s=\frac{i}{\sqrt{2}}d_s+\frac{1}{\sqrt{2}}a_s,& b_s=\frac{1}{\sqrt{2}}d_s+\frac{i}{\sqrt{2}}a_s,\label{CTC0}
\end{eqnarray}         
and from the properties of the CTC we must have 
\begin{eqnarray}
d_s=r_{tot.}(E,E')r'_{tot.}(E,E')e^{i\xi}\sum_{s'}U_{s,s'}c_{s'}=-r^2_{tot.}(E,E')e^{i\xi}\sum_{s'}U_{s,s'}c_{s'}\label{CTC0B}
\end{eqnarray} with $\xi=\chi-2E(t_\mathcal{B}-t_\mathcal{A})$ and where $U_{s,s'}$ are matrix elements defined as $U_{s,s'}=\psi^\dagger_s(\mathbf{p},E)\hat{U}\psi_{s'}(\mathbf{p},E)$. 
Writing the coefficients $a_s$, $b_c$, etc.. as two components vectors $\underline{a}:= \left(\begin{array}{c}a_{+\frac{1}{2}}\\ a_{-\frac{1}{2}}
  \end{array}\right)$, $\underline{b}:= \left(\begin{array}{c}b_{+\frac{1}{2}}\\ b_{-\frac{1}{2}}
  \end{array}\right)$, etc... and introducing the unitary matrix $\underline{U}:= \left(\begin{array}{cc}U_{+\frac{1}{2},+\frac{1}{2}}& U_{+\frac{1}{2},-\frac{1}{2}} \\ U_{-\frac{1}{2},+\frac{1}{2}}& U_{-\frac{1}{2},-\frac{1}{2}}\end{array}\right)$ we can solve Eqs.~\ref{CTC0}, \ref{CTC0B} and we obtain:
\begin{eqnarray}
\underline{c}=\frac{1}{\sqrt{2}}\frac{1}{\underline{I}+\frac{ir^2_{tot.}(E,E')}{\sqrt{2}}e^{i\xi}\underline{U}}\underline{a},\nonumber\\
\underline{b}=\frac{i}{\sqrt{2}}[\underline{I}+\frac{ir^2_{tot.}(E,E')}{\sqrt{2}}e^{i\xi}\underline{U}\frac{1}{1+\frac{ir^2_{tot.}(E,E')}{\sqrt{2}}e^{i\xi}\underline{U}}]\underline{a}.\label{CTC1}
\end{eqnarray}  In QED   taking into account $P_v$ Eq.~\ref{CTC1} gives the probability for detecting a single electron in $D_1$ assuming no other detection in $D_{2,3}$:
\begin{eqnarray}
P(D_1,\varnothing D_{2,3})=P_v\lVert \underline{b}\rVert^2=\frac{P_v}{2}\lVert [\underline{I}+\frac{ir^2_{tot.}(E,E')}{\sqrt{2}}e^{i\xi}\underline{U}\frac{1}{1+\frac{ir^2_{tot.}(E,E')}{\sqrt{2}}e^{i\xi}\underline{U}}]\underline{a}\rVert^2.
\end{eqnarray}  where by definition $\lVert \underline{b}\rVert^2:=\underline{b}^\dagger\underline{b}:=\sum_s b_s^\ast b_s$.  Similarly we obtain  $P(D_3,\varnothing D_{1,2})=\frac{P_vT_{tot.}(E,E')}{2}\lVert{ \frac{1}{\underline{I}+\frac{ir^2_{tot.}(E,E')}{\sqrt{2}}e^{i\xi}\underline{U}}\underline{a}}\rVert^2$.  Computing $P_v$ in this process is a bit more complicated than in the previous cases.   We need to consider single pair production in 3 channels A) an antiparticle is coming from $D_2$ and ends in $D_3$, B) an antiparticle is coming from $D_2$ and ends in $D_1$ or C) an antiparticle is coming from $D_3$ and ends in $D_2$ (in this channel the CTC is used by an antielectron going in the opposite, i.e., anticlockwise direction).    We must also include double-pairs production obtained: D) by combining A) with C),  and  E) combining B) with C). We have: 
\begin{eqnarray}
P_A=P_v R_{tot.}(E,E')\textrm{Tr}[\underline{M}_A^\dagger \underline{M}_A],& P_B=\frac{P_v}{2} T_{tot.}(E,E')\textrm{Tr}[\underline{M}_B^\dagger \underline{M}_B]\nonumber\\
P_C=P_v R_{tot.}(E,E')\textrm{Tr}[\underline{M}_C^\dagger \underline{M}_C],&
P_D=P_v R^2_{tot.}(E,E')\textrm{Tr}[\underline{M}_A^\dagger \underline{M}_A]\textrm{Tr}[\underline{M}_C^\dagger \underline{M}_C]\nonumber\\P_E=\frac{P_v}{2}R_{tot.}(E,E') T_{tot.}(E,E')\textrm{Tr}[\underline{M}_B^\dagger \underline{M}_B]\textrm{Tr}[\underline{M}_C^\dagger \underline{M}_C]
\end{eqnarray} with 
\begin{eqnarray}
\underline{M}_A= \underline{I}+\frac{it^2_{tot.}(E,E')}{\sqrt{2}}e^{i\xi}\underline{U}\frac{1}{1+\frac{it^2_{tot.}(E,E')}{\sqrt{2}}e^{i\xi}\underline{U}}\nonumber\\
\underline{M}_B= \underline{U}\frac{1}{1+\frac{it^2_{tot.}(E,E')}{\sqrt{2}}e^{i\xi}\underline{U}}\nonumber\\
\underline{M}_C= \underline{I}-\frac{i\sqrt{\alpha} t^2_{tot.}(E,E')}{\sqrt{2}}e^{i\xi}\underline{V}\frac{1}{1-\frac{i\sqrt{\alpha} t^2_{tot.}(E,E')}{\sqrt{2}}e^{i\xi}\underline{V}}
\end{eqnarray} where $V_{s,s'}=\psi^\dagger_s(-\mathbf{p},-E)\hat{U}^{-1}\psi_{s'}(-\mathbf{p},-E)=\psi^\dagger_{s'}(-\mathbf{p},-E)\hat{U}\psi_{s}(-\mathbf{p},-E)$ and $\alpha$ is a correcting coefficient for the reflectivity of $BS$ since this device is not in general a balanced 50/50 beam splitter for negative energy electrons. In the end we obtain the sum rule:
\begin{eqnarray}
1=P_v+P_A+P_B+P_C+P_D+P_E
\end{eqnarray} that allows us to compute $P_v$.\\
\indent 2) In the second strategy, illustrated in Fig.~\ref{fig:12}(b), the single electron after interacting with a first 50/50 beam splitter $BS_1$ enters the CTC and interfere with it self at the second beam splitter $BS_2$ with the second beaam that was reflected by $BS_1$. The amplitudes $a$ and $d$ at the entrance  of $BS_2$ interacts and this correspond to the `grand father paradox' as defined by Deutsch \cite{Deutsch} where we an kill you parent in a parallel universe.    In particular, we can tune the phase delay in the full interferometer involving a unitary   $\hat{U}$ in order to cancel the amplitude $c_s$ in the $c-$channel:
\begin{eqnarray}
\underline{c}=\frac{i}{2}(\underline{I}e^{i\chi}e^{-iE(t_{BS_2}-t_{BS_2})}-r^2_{tot.}(E,E')e^{i\xi}\underline{U})\underline{a}=0
\end{eqnarray} 
where as before $\xi=\chi-2E(t_\mathcal{B}-t_\mathcal{A})$ and where a delay associated with the  two interactions at $BS_1$ and $BS_2$ has been included. This equation is easy to solve if  $\hat{U}=1$ and we obtain the only condition 
\begin{eqnarray}
r^2_{tot.}(E,E')e^{i(\xi-\chi+E(t_{BS_2}-t_{BS_2})}=1
\end{eqnarray} that always admits a solution (by tuning the phases) if $R_{tot.}(E,E')=1$, a regime that can be reached in the high electromagnetic field regime (see Appendix~\ref{Sec2}).
In the configuration where $\hat{U}=1$ we can define $P_v$ using the formula
\begin{eqnarray}
1=P_v[1+T_{tot.}(E,E')R_{tot.}(E,E')(1+\alpha)+\alpha T^2_{tot.}(E,E')R^2_{tot.}(E,E')]
\end{eqnarray}
Moreover, we have in the $d-$channel 
\begin{eqnarray}
P(D_1,\varnothing D_{2,3,4})=P_v\lVert \underline{b}\rVert^2=\frac{P_v}{2}|e^{i\chi}e^{-iE(t_{BS_2}-t_{BS_2})}+r^2_{tot.}(E,E')e^{i\xi}|^2.
\end{eqnarray} In the `Deutschian-grand father' regime $R_{tot.}(E,E')=1$ (and thus $T_{tot.}(E,E')=2$) used here this probability reduces to:
 \begin{eqnarray}
P(D_1,\varnothing D_{2,3,4})=2P_v=\frac{2}{3+6\alpha}\leq \frac{2}{3}.
\end{eqnarray} 
\indent In the end we mention (see Fig.~\ref{fig:13}) that a gravitational wormhole could be used for implementing QS. Paths of a single particle  could made loops passing through interacting regions $A$ and $B$ located outside Fig.~\ref{fig:13}(a) of inside the time tunnel Fig.~\ref{fig:13}(b). The interference between different paths leads to indefinite causal order   but with a fixed time order structure. This clearly motivated the present work in order to find equivalents  of such speculative wormholes with a relativistic electron scattered by two or more temporal Fabry-Perot cavities.       
\indent Other interesting implementations of the quantum switch are clearly possible by using more than two cavities and CTCs.  For example, we show in Fig.~\ref{fig:14}(a) the simplest QS based on a single beam splitter for photon   (a versioon using a polarising beam spliter is also possible).   Here the target is the polarisation of the single photon in this interferometer  passing through two counter propagating paths  involving two unitaries  $\hat{U}_{A,B}$.      The same circuit is possible with a relativistic electron making a CTC          between three temporal cavities as illustrated in Fig.~\ref{fig:14}(b).The analysis  of this quantum switch can be done with the techniques developped in this work.  


\begin{thebibliography}{}
\bibitem{remark}  $\psi^{(2)}(x)$ also includes $\int_{\Omega_{\mathcal{A}}}\int_{\Omega_{\mathcal{A}}} d^4x_{\mathcal{A}}d^4x'_{\mathcal{A}}\chi(x,x_{\mathcal{A}}, x'_{\mathcal{A}})+\int_{\Omega_{\mathcal{B}}}\int_{\Omega_{\mathcal{B}}} d^4x_{\mathcal{B}}d^4x'_{\mathcal{B}}\chi(x,x_{\mathcal{B}}, x'_{\mathcal{B}})$ that we ommited here.
\bibitem{notev} The mode volume $V$ is associated with the finite spatial extension of a quasi manochromatic wavepacket.
\end{thebibliography}
\end{document}